\documentclass[pdftex,12pt]{iopart}
\usepackage{graphicx}
\usepackage{float}
\usepackage{ulem}
\usepackage[dvips]{color}
\begin{document}

\title{Quantum resistance metrology using graphene}

\author{T.J.B.M. Janssen$^1$, A. Tzalenchuk$^{1,2}$, S. Lara-Avila$^3$, S.~Kubatkin$^3$, and V.I.~Fal'ko$^4$}

\address{$^1$National Physical Laboratory, Hampton Road, Teddington TW11 0LW, UK}
\address{$^2$Royal Holloway, University of London, Egham Hill, Egham, TW20 0EX, UK}
\address{$^3$Department of Microtechnology and Nanoscience, Chalmers University of Technology, S-412 96 G\"{o}tenborg, Sweden}
\address{$^4$Physics Department, Lancaster University, Lancaster LA1 4YB, UK}

\ead{jt.janssen@npl.co.uk}

\begin{abstract}
In this paper we review the recent extraordinary progress in the development of a new quantum standard for resistance based on graphene. We discuss the unique properties of this material system relating to resistance metrology and discuss results of the recent highest-ever precision direct comparison of the Hall resistance between graphene and traditional GaAs. We mainly focus our review on graphene expitaxially grown on SiC, a system which so far resulted in the best results. We also briefly discuss progress in the two other graphene material systems, exfoliated graphene and chemical vapour deposition graphene, and make a critical comparison with SiC graphene. Finally we discuss other possible applications of graphene in metrology.   
\end{abstract}

\maketitle
\tableofcontents

\section{Introduction}
The discovery eight years ago of the quantum Hall effect (QHE) in graphene sparked an immediate interest in the metrological community. The QHE is a fascinating macroscopic quantum effect occurring in two-dimensional conductors and relates the resistance quantum, $h/e^2$ only to the fundamental constants of nature, $h$, the Planck constant and, $e$, the elementary charge \cite{Novoselov2005,Zhang2005}. Although metrology has successfully used the QHE for more than two decades to realise the resistance scale \cite{Jeckelmann2001}, graphene is a material with properties like no other. Graphene, a single layer of carbon atoms in a hexagonal crystal lattice structure, is a truly two-dimensional metal with a linear dispersion relationship characteristic for massless Dirac-type charge carriers \cite{Geim2009}. The unique bandstructure of this semi-metal has both practical and fundamental implications. Firstly, the massless nature of the charge carriers leads to a Landau level spectrum with an energy gap between the first two levels which is around five times larger than that in semiconductor materials for magnetic fields around 10 tesla. This implies that the QHE in graphene can be observed at much reduced magnetic fields and/or much higher temperatures \cite{Novoselov2007}. Secondly, the marked difference in bandstructure and charge carrier characteristics between graphene and semiconductor systems allows for a demonstration of the universality of the quantum Hall effect through a rigorous test of the material independence of the value of $R_{\rm K}=h/e^2$, the von Klitzing constant.   

Theory predicts no major corrections to the simple relation $R_{\rm K}=h/e^2$. The quantum Hall resistance is considered to be a topological invariant~\cite{Thouless1994}, not altered by the electron-electron interaction, spin-orbit coupling, or hyperfine interaction with the nuclei. It has also been shown that the quantized Hall resistance is insensitive to much more subtle influences of the gravitational field \cite{Hehl2004}. Recently, a quantum electrodymical approach to charge carriers in a magnetic field has predicted a tiny correction to the von Klitzing constant of the order of $10^{-20}$ for practical magnetic field values~\cite{Penin2009}. However, the size of this predicted correction is about 8 to 10 orders of magnitude smaller than the most accurate measurement techniques available and therefore untestable. Nevertheless, the fundamental nature of the Hall resistance quantization makes experimental tests of its universality of the utmost importance. Universality of $h/e^2$ will also strongly support the pending redefinition of the SI-units for kilogram and ampere in terms of $h$ and $e$~\cite{Mills2006}.

\begin{figure*}
\includegraphics[scale=0.6]{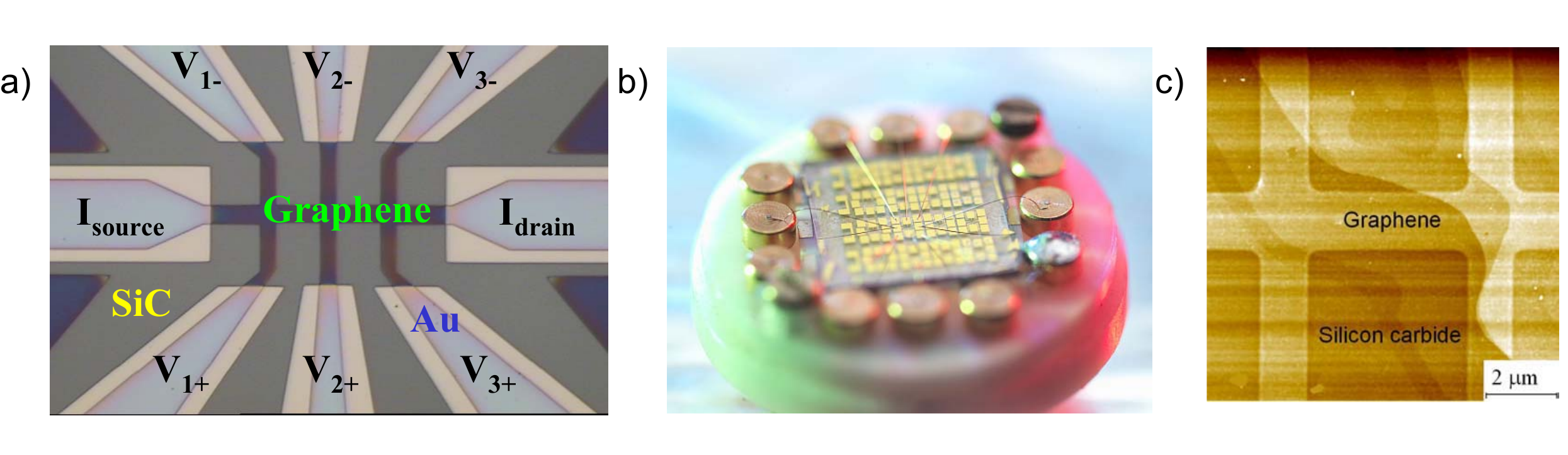}
\caption{\label{labelfig1}(Color online) Graphene on SiC device used for high-precision quantum Hall resistance measurements. (a) Optical micrograph of a Hall bar. (b) Layout of a $7\times 7\ \rm mm^2$ wafer with 20 Hall bars. (c) AFM image of Hall bar. Meandering lines are steps in the surface. (Adapted from Ref.~\cite{Tzalenchuk2011})}
\end{figure*}

A comparison of the Hall resistance in two different substances does not prove the exactness of the relationship $R_{\rm K}=h/e^2$, however, material independence is a significant factor in establishing the fundamental nature of $R_{\rm K}$. This material independence turns out to be rather difficult to establish. Indeed the characteristics of QHE samples must satisfy very stringent requirements~\cite{Delahaye2003} and in 30 years only silicon MOSFETs (metal-oxide-semiconductor field-effect transistors) and III-V semiconductors (GaAs/AlGaAs or InGaAs/InP heterostuctures) did so~\cite{Jeckelmann2001}. 

The first accurate measurements of the QHE in graphene were performed by Giesbers {\it et al.}~\cite{Giesbers2008} on exfoliated samples. The precision obtained in these measurements was at the part-per-million (ppm) level and limited by the high contact resistances together with a low maximum source-drain current which these samples could sustain before breakdown of the QHE occured. A large measurement current determines the maximum signal-to-noise ratio and increasing this breakdown current is key to high-accuracy measurements. One established method of increasing the breakdown current is to increase the sample width~\cite{Jeckelmann2001} which is not easy to achieve with the exfoliation technique. 

A breakthrough came in 2009, when several groups within days of each other succeeded in growing large-area wafers of epitaxial graphene by sublimation of SiC (SiC/G) with a quality good enough to observe quantized Hall resistance~\cite{Shen2009, Wu2009, Tzalenchuk2010, Jobst2010, Tanabe2010}. In an indirect comparison with a GaAs/AlGaAs heterostructure device via an intermediate room-temperature standard resistor, Tzalenchuk {\it et al.}~\cite{Tzalenchuk2010} were the first to demonstrate an accuracy of 3 parts in $10^9$ (ppb) for the resistance quantization in a large SiC/G device (see fig.~\ref{labelfig1}). The measurement system~\cite{Williams2010} was of identical design to that used by Giesbers {\it et al.}~\cite{Giesbers2008} and the key factor in the large improvement was the very low contact resistances which could be achieved and an order of magnitude increase in the breakdown current. The large breakdown current in SiC/G is not simply the result of a larger device but finds its origin in a charge exchange mechanism between the SiC substrate and graphene, leading to an unusually strong pinning of the  quantum Hall state~\cite{Kopylov2010,Janssen2011a}. This mechanism turns out to be stronger for lower carrier densities and by using a novel photochemical gating technique, Lara-Avila {\it et al.}~\cite{Lara-Avila2011} have been able to develop a graphene device with an extremely robust $\nu=2$ quantum Hall state. Robustness in this context means invariance of the quantization under changes of magnetic field, temperature or source-drain current and is a key requirement for any practical application. Recently, the same team undertook a direct comparison between SiC/G and GaAs, using an enhanced measurement system, demonstrating equivalence of $R_{\rm H}$ with a relative uncertainty of 8.7 parts in $10^{11}$~\cite{Janssen2011b}.

\section{Quantum Hall effect in conventional 2DEGs}

\begin{figure}[t]
  \centering
 {\label{labelfig2}\includegraphics[scale=.85]{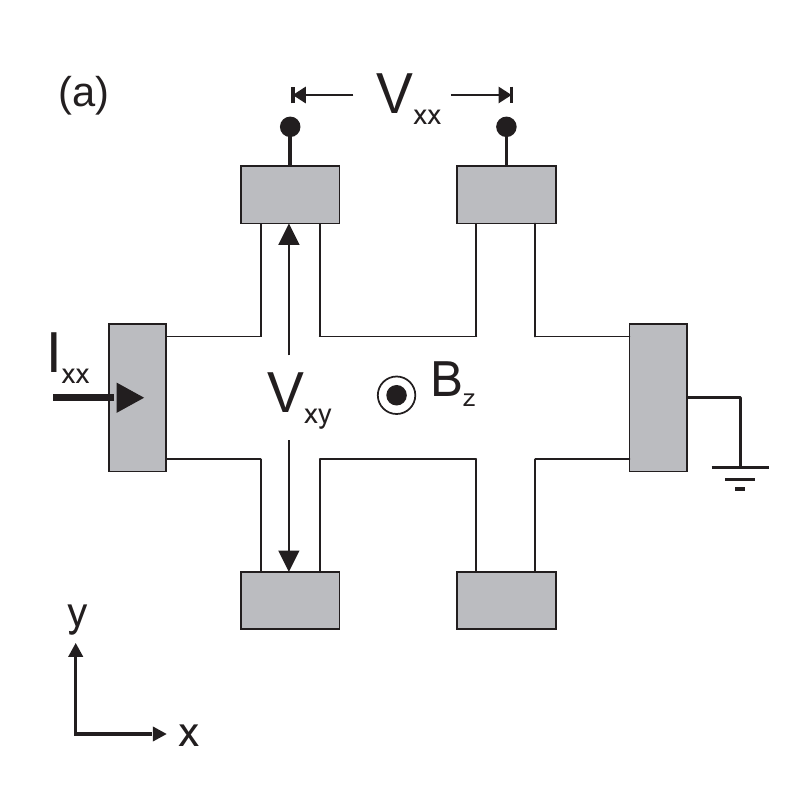}}
 {\includegraphics[scale=.85]{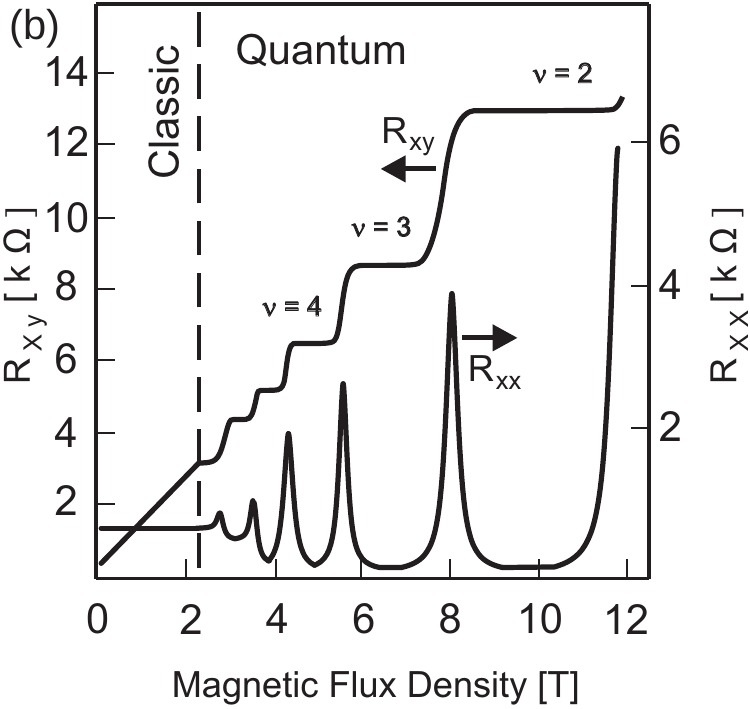}}  
\caption{(a) Magnetotransport measurements in a Hall bar geometry enable the electrical characterization of a material in terms of carrier type (electrons or holes), concentration and mobility. (b) Classic (quantum) Hall effect at low (high) magnetic fields in conventional, semiconductor-based, two-dimensional systems.} 
\end{figure}

When a conducting material is placed simultaneously in electric ($\vec{E}$) and magnetic ($\vec{B}$) fields, charge carriers ($e$) (electrons or holes) moving with a velocity $\vec{v}$ experience a magnetic (Lorentz) force proportional to $\pm e\left( \vec{E}+\vec{v} \times \vec{B} \right)$. In the Hall bar geometry of fig. \ref{labelfig2}a, carriers moving in the $x$-direction ($I_{xx}$) will be deflected by the perpendicular magnetic field $B_z$ and accumulated at the edges of the sample. This accumulation of carriers gives rise to a potential build up in the direction transversal to carrier propagation ($V_{xy}$), the Hall voltage. Under these conditions the resistivity\footnote{For a two-dimensional conductor the 2D resistivity is defined as $\rho_{2D}=\rho_{3D} /t$, with $t$ the (infinitesimal) thickness of the sample. $\rho_{2D}$ has the same dimensions as the 3D resistance ($\Omega $). In order to avoid confusion, the units of $\rho_{2D}$ are most of the time explicitly written as $\Omega /(L/W)=\Omega$/square. Sometimes, however, the words \textit{resistance} and \textit{resistivity} for two-dimensional samples are used indistinctly.} of the sample takes a tensor form, being different in the direction parallel or perpendicular to the direction of the current. We define the \textit{transversal} resistivity $\rho_{xy}$ as:

\begin{equation}\label{labeleq1}
       \rho_{xy}= \frac{V_{xy}}{I_{xx}}= \frac{B_z}{n_se},
\end{equation}

where $n_s$ is the sheet carrier density. The constant of proportionality between the transversal resistivity and the applied magnetic field is called the \textit{Hall} coefficient ($r_H=1/n_se$). The \textit{longitudinal} resistivity of the sample $\rho_{xx}$, on the other hand, is independent of magnetic field and can be obtained by measuring the voltage developed along the direction of the current and taking into account the geometry of the sample (length $L$ and width $W$).

\begin{equation}\label{labeleq2}
       \rho_{xx}=\frac{W}{L} \frac{V_{xx}}{I_{xx}}
\end{equation}

In general, low-field Hall effect measurements are a simple yet powerful tool to study electronic properties of materials. The magnitude and sign of the Hall coefficient have been extensively used to characterize electronic materials in terms of carrier type (electrons or holes) and density $n_s$ (per unit area). In addition, the Hall coefficient and the longitudinal resistivity can be combined to find the average carrier mobility in the sample as  $\mu=r_{\rm H}/\rho_{xy}$.

The \textit{classic} Hall effect picture breaks down when working with high mobility, two-dimensional systems at low temperatures and in strong magnetic fields \cite{vonKlitzing1980} . Under these conditions, quantum effects are manifest as quantized steps in $\rho_{xy}$, while  simultaneously the longitudinal resistance vanishes, $\rho_{xx}=0$ (fig. \ref{labelfig2}b). Steps and plateaux in the transversal resistance occur at universal values of $\rho_{xy}=h/\nu e^2$, with $h$ the Planck constant, $e$ the elementary charge and $\nu$ and integer. This is the quantum Hall effect ~\cite{vonKlitzing1980}.

The QHE can be understood as a consequence of Landau quantization \cite{Landau1977}. Under the presence of a strong magnetic field, the size of the cyclotron orbit shrinks and becomes comparable to the wavelength of carriers. If the mobility in the system is high enough so that a carrier can complete a few cyclotron orbits before its momentum is relaxed, that is:

\begin{equation}\label{labeleq3}
       \omega_c^{-1} \ll \tau
\end{equation}

then the size of the orbit can only take some allowed values, corresponding to an integer number times the electron wavelength ($\tau$ is the relaxation time). Recalling that the cyclotron frequency is given by $\omega_c=eB/m^*$ and the carrier mobility can be expressed as $\mu=e\tau / m^*$, condition (\ref{labeleq3}) can be rewritten as $ B \gg \mu^{-1} $ ($m^*$ is the effective mass). Thus, quantum effects can in principle be observed with $B=1$~T if the mobility of the sample is around $\mu=10,000$ cm$^2$V$^{-1}$s$^{-1}$. 

\begin{figure}[t]
  \centering{\includegraphics[scale=.85]{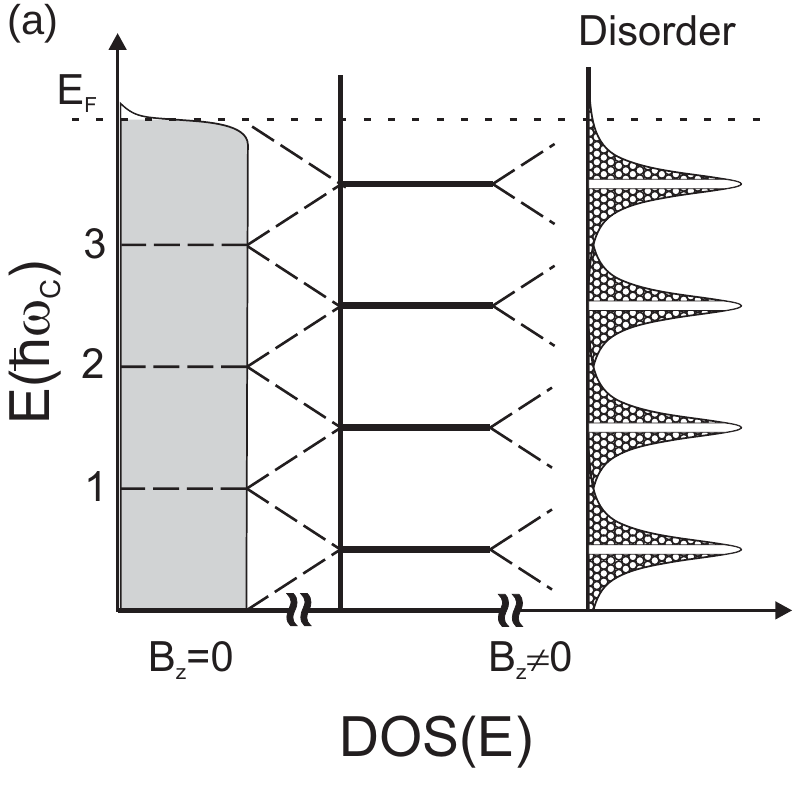}}
  {\includegraphics[scale=.85]{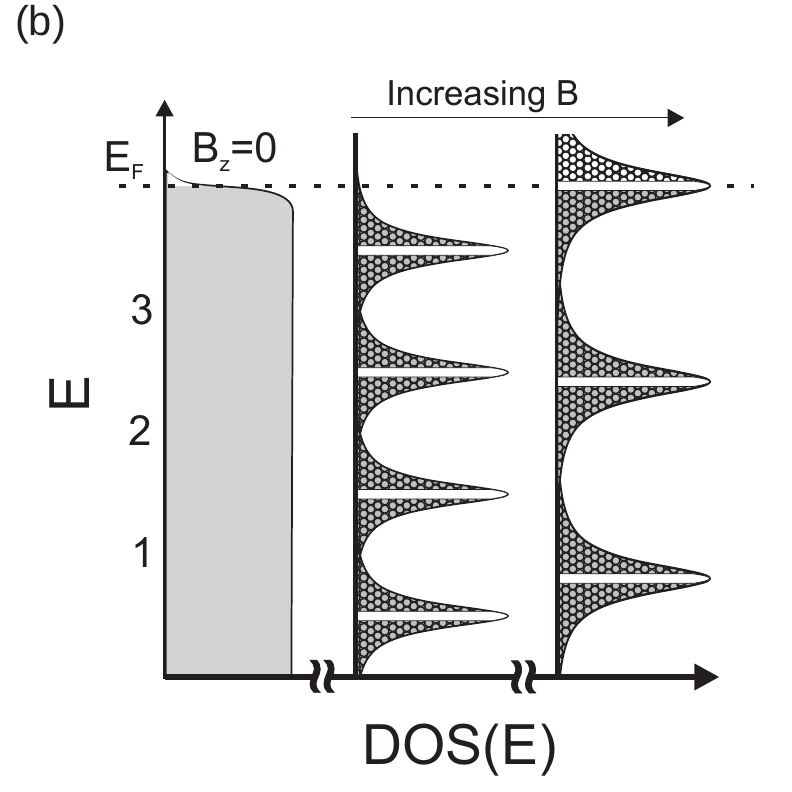}}
  \caption{(a) Quantization of cyclotron motion leads to formation of Landau levels (LL) in the density of states (DOS). In theory the levels, spaced by $\hbar \omega_c$, are highly degenerate delta functions, but in real samples the presence of disorder broadens the levels. (b) In the absence of electrostatic gate the electron concentration remains constant and the quantum Hall plateaux are observed by increasing the magnetic field $B$, whenever the Fermi energy lies in between the center of Landau levels.}\label{labelfig3}
\end{figure}

Quantized cyclotron motion modifies dramatically the density of states by breaking it into discrete levels, so-called Landau levels (LL), which are the allowed energies for cyclotron orbits under quantizing conditions (fig. \ref{labelfig3}a). Formally, LL are obtained by solving the Schr\"odinger equation for free electrons in a 2D systems, in the presence of electric and magnetic fields. The problem reduces to that of a harmonic oscillator shifted by the magnetic length $\ell_B= \sqrt{\hbar/eB}$, with eigenvalues given by \cite{Landau1930,Landau1977,QHE90}:
\begin{equation}\label{labeleq4}
      E_{N}= \hbar\omega_c (N+\frac{1}{2})
\end{equation}
with the reduced Planck constant $\hbar=h/2\pi$, the cyclotron frequency $\omega_c=eB/m^*$, and $N$ an integer, zero included. In theory, each LL is a heavily degenerate delta function in the density of states filled with localized states undergoing cyclotron motion. In real samples, nevertheless, LL are broadened by the presence of disorder, (fig. \ref{labelfig3}a) due inelastic collisions (energy exchange) of localized electrons with impurities. 

The steps observed in $\rho_{xy}$ in the QHE can be explained by considering that LL are filled with localized carriers undergoing cyclotron motion. As carriers undergo cyclotron motion, they enclose a quantum of magnetic flux $\Phi_0=h/e$,  and thus the number of localized carriers per unit area, $n_{LL}$, in the system can be found as:
\begin{equation}\label{labeleq5}
      n_{LL}=B/\Phi_0=eB/h
\end{equation}

With $n_{LL}$ known, the number of completely filled Landau levels $\nu$ (so-called filling factor) can be found by dividing the total density of electrons in the system, $n_s$, by the number of localized carriers, $n_{LL}$. 
\begin{equation}\label{CH1:filling_factor}
      \nu=n_s/n_{LL}
\end{equation}

When a LL is full the Fermi level lies in a gap between occupied levels and the filling factor $\nu$ in Eq.~\ref{CH1:filling_factor} must be an integer (fig. \ref{labelfig3}a). If we substitute the discrete density of states resulting from the formation of LL into the field-dependence of $\rho_{xy}$ for the classic Hall effect (Eq. \ref{labeleq1}), we find that $\rho_{xy}$ is quantized as:

\begin{equation}\label{labeleq6}
      \rho_{xy}=\frac{B}{en_s}=\frac{B}{e\nu n_{LL}}=\frac{h}{\nu e^2 }
\end{equation}

A common interpretation for the QHE is in the picture of extended states, carrying current without dissipation (zero resistance) along the edges of the sample, and localized states undergoing cyclotron motion in the bulk \cite{Halperin1982,Buttiker1988}. The origin of zero longitudinal resistance $\rho_{xx}$ is that extended states propagating in one direction of the sample are spatially separated from those carrying current in the opposite direction, thereby suppressing backscattering. Maxima in $\rho_{xx}$ are observed every time the Fermi level crosses the center of a LL; plateaux in $\rho_{xy}$ and vanishing $\rho_{xx}$ are observed whenever the Fermi level lies in between the center of LL, pinned by localized states. Experimentally,  quantum Hall plateaux can be observed by: a) fixing the magnetic field and varying the Fermi level of the sample with e.g. an electrostatic gate or b) fixing the Fermi level (fixed carrier concentration) and varying the magnetic field (fig. \ref{labelfig3}b).

\section{Resistance metrology}\label{resistance-metrology}

\subsection{Requirements}\label{Requirements}
In resistance metrology the value of the quantized Hall resistance, either at $R_{\rm H}\sim 12.9\ \rm k\Omega$\footnote{Here $R_{\rm H}$ refers to the quantum Hall resistance, defined as $R_{\rm H}=V_{xy}/I_{xx}$, as opposed to the Hall coefficient $r_{\rm H}=1/n_se$ mentioned earlier} for $\nu=2$ or at $\sim 6.4\rm \ k\Omega$ for $\nu=4$, needs to be compared with standard resistors at decade values typically in the range of 1~$\Omega$ to 100~$\rm k\Omega$. Room temperature measurement systems such as potentiometric or current transformer bridges can perform these measurements to a level of a few parts in $10^8$ \cite{Jeckelmann2001}. However, most top-level national measurement laboratories make precision measurements with a so-called cryogenic current comparator (CCC) bridge~\cite{Williams2011b}. A CCC bridge can be used to determine the ratio of almost any pair of resistors with an accuracy of a few parts per billion (ppb)~\cite{Williams2010}. If one of these resistors is a quantum Hall device, it can be used to realise the value of the other resistor in terms of the SI ohm via the internationally agreed value of the von Klitzing constant, $R_{\rm K-90}$~\cite{Petley92}. If both resistors are quantum Hall devices, the CCC bridge can be employed to determine the universality of the quantum Hall effect~\cite{Hartland1991}. 

\begin{figure}
\centering
\includegraphics[scale=0.6,angle=-90]{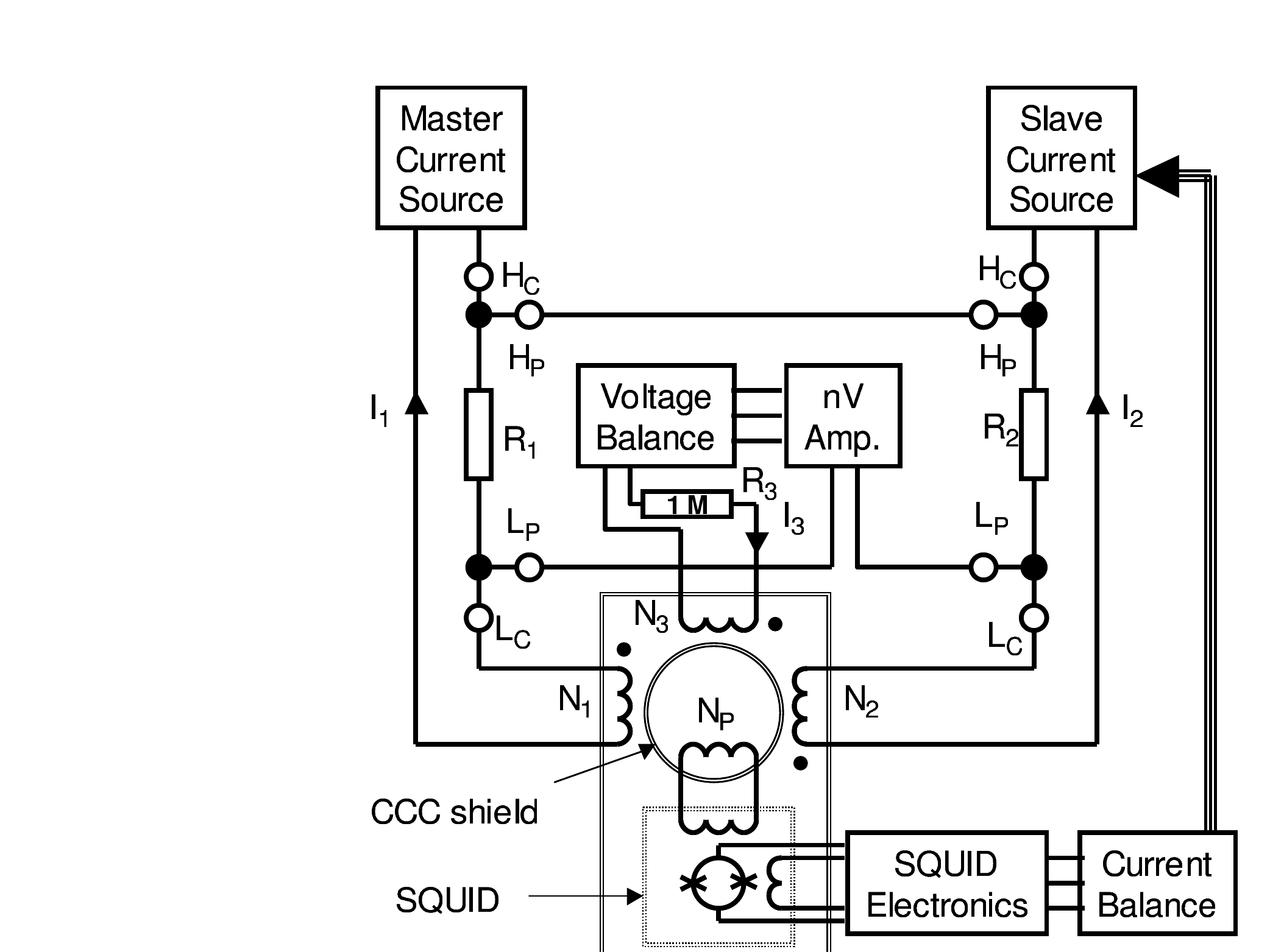}
\vspace{1cm}
\caption{\label{labelfig4} Simplified schematic of the cryogenic current comparator bridge circuit. (Adapted from Ref.~\cite{Williams2010})}
\end{figure}

The basic principle of a CCC can be understood by considering a simple superconducting tube with two current-carrying wires running through. Magnetic flux generated by the wires is excluded from the superconductor by the Meissner effect through a screening current on the surface of the superconductor. The flux generated by the screening current on the outside of the tube is sensed by a pick-up coil connected to a SQUID (Superconducting Quantum Interference Device). When the currents in the two wires are equal and opposite the screening current is exactly zero and, importantly, independent of the position of the wires inside the tube which allows for the very high accuracy of this device.

The bridge circuit for measuring the ratio of two resistors in terms of the current ratio determined by a CCC is schematically shown fig.~\ref{labelfig4}. Two isolated current sources ``Master'' and ``Slave'' separately drive current through resistors $R_1$ and $R_2$ and associated windings $N_1$ and $N_2$ on the CCC. The current ratio can be set via electronics to a few parts in $10^6$ and this ratio is improved to a level of 1 part in $10^{11}$ by forming a negative feedback loop from the SQUID sensing the net flux in the CCC to the Slave current source. When balanced, the reading on the nanovoltmeter is exactly proportional to the resistance ratio (often a second servo loop is used to null the nanovoltmeter reading by injecting a current in a third, $N_3$, small winding on the CCC~\cite{Williams2010}). The main uncertainty components of the bridge are, the noise of the SQUID sensor, the Johnson noise of the resistors being measured and the current and voltage noise of the null detector. These components combine to give a typical measurement uncertainty of 1 ppb for a measurement time between 5 and 15 minutes~\cite{Williams2010}.

In principle the value of the quantized Hall resistance is exactly $h/\nu e^2$ when the magnetic field is set to integer filling factor $\nu$. However, in practice there are many possible effects which can cause the Hall resistance to deviate from exact quantization. One of the key factors is unwanted dissipation: if $\rho_{xx}$ is not exactly zero, the  Hall device is not in the non-dissipative state and $\rho_{xy}$ will deviate from exact quantization. Also if the contact resistances to the 2DEG are too large (typically $\geq 10\ \Omega$), excessive dissipation will occur at the contacts which again can cause errors in $R_{\rm H}$. Over the last two decades the metrological community has evaluated all possible errors in the quantum Hall effect and composed a set of experimental guidelines which, when precisely followed, will result in a measurement uncertainty smaller than 1~ppb \cite{Jeckelmann2001,Delahaye2003}.

\subsection{State-of-the-art}
\begin{figure}
\centering
\includegraphics[scale=0.8]{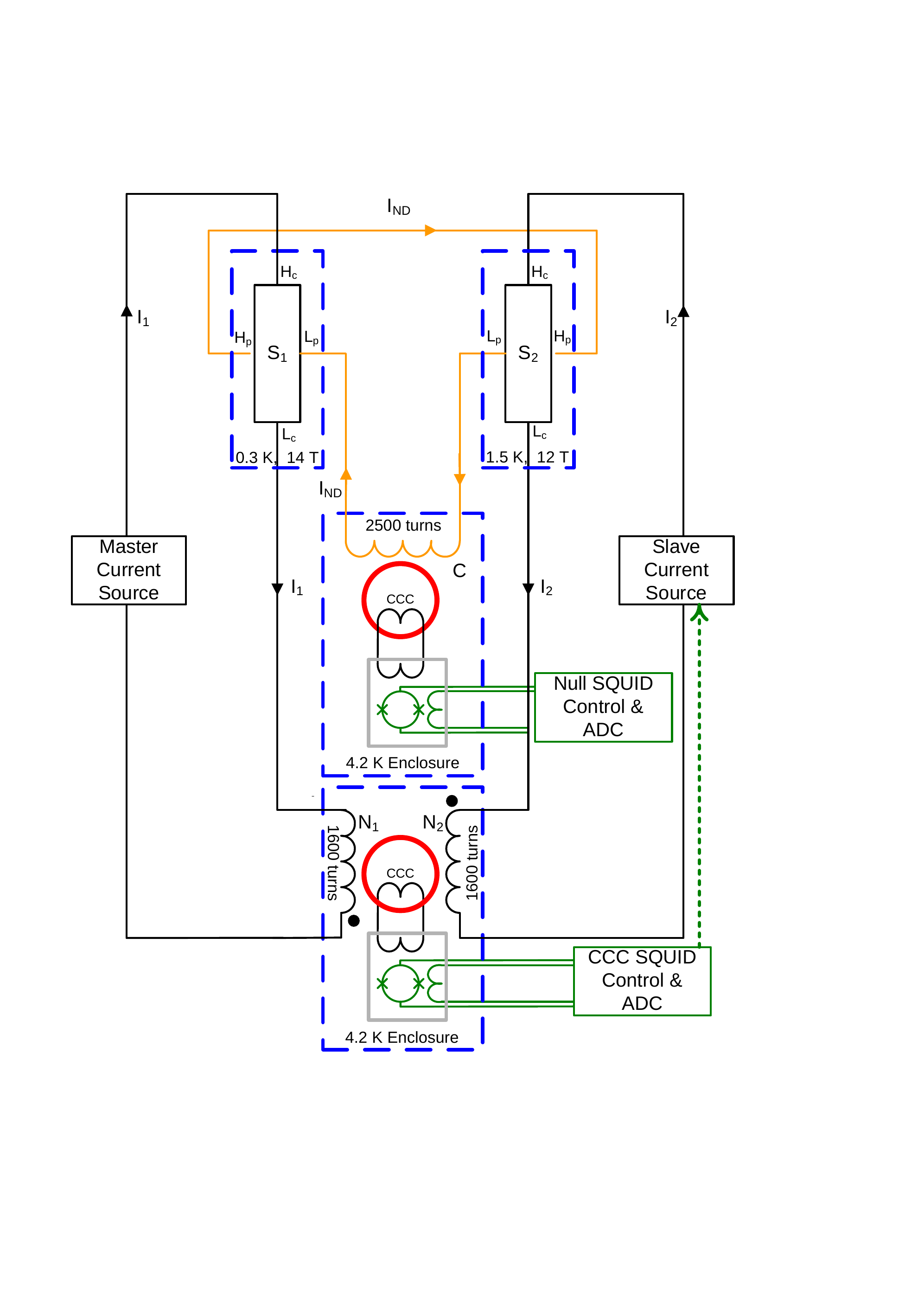}
\vspace{-3cm}\caption{\label{labelfig5} (Color online) Schematic of a cryogenic current comparator bridge circuit used for high precision universality test. (Adapted from Ref.~\cite{Janssen2011b,Janssen2012})}
\end{figure}

When measuring two quantum Hall devices with a combined impedance of $\approx 25\ \rm k\Omega$ the current-noise properties of the nanovoltmeter start to dominate the uncertainty budget. This problem can be alleviated by employing a second CCC in place of the nanovoltmeter (see fig.~\ref{labelfig5} where $R_1$ and $R_2$ are replaced by QHR samples $S_1$ and $S_2$). The potential contacts on $S_1$ and $S_2$ are closed in a loop via winding C on this second CCC (see fig.~\ref{labelfig5}). This device is configured with just a single winding to measure a current null rather than two windings to establish a current ratio. Janssen {\it et al.}~\cite{Janssen2012} made a careful evaluation of all uncertainty components of this measurement system and demonstrated that a systematic uncertainty of only 1 part in $10^{11}$ can be achieved for a direct 1:1 comparison of two quantum Hall devices.

\begin{figure}
\centering
\includegraphics[scale=0.8]{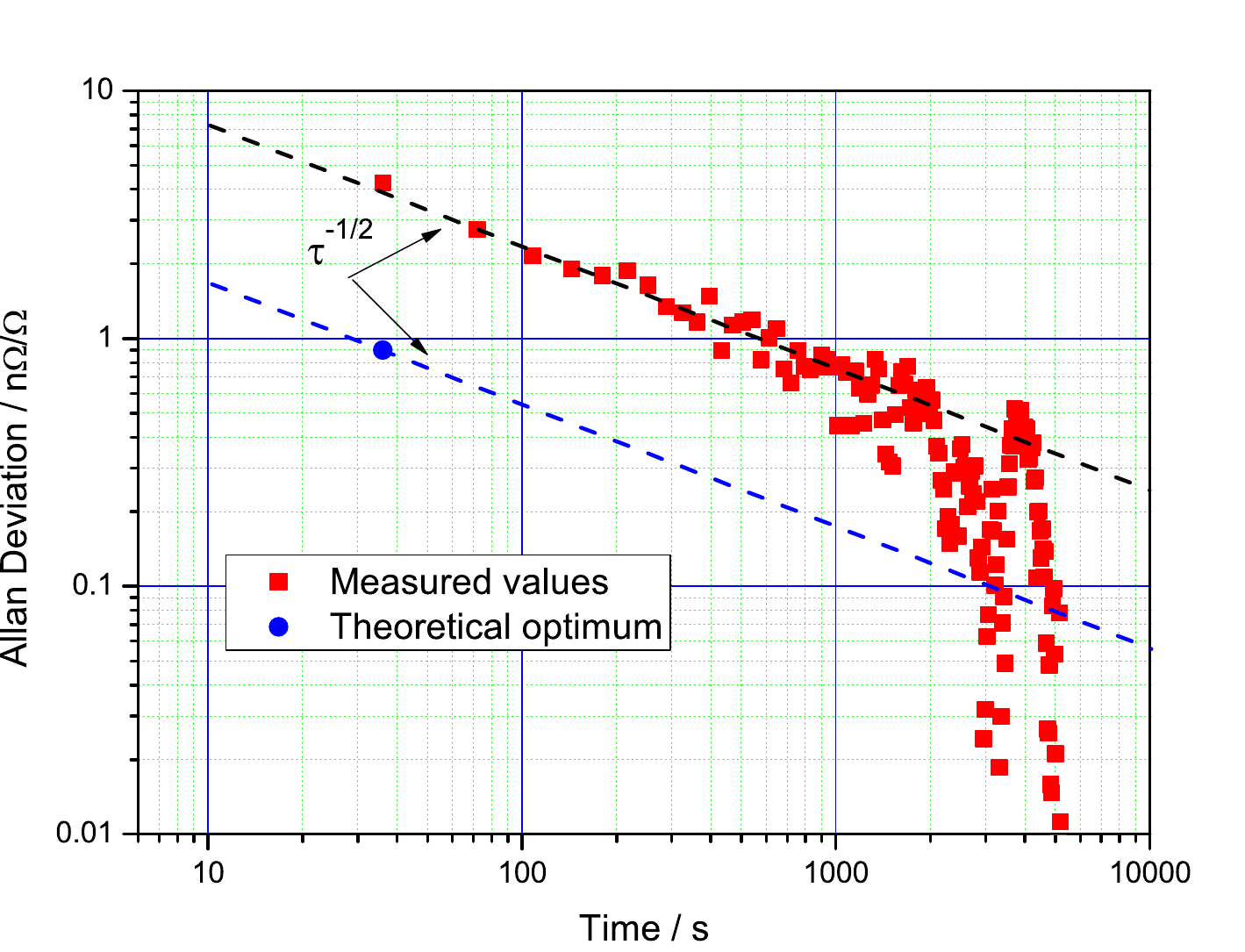}
\caption{\label{labelfig6} (Color online) Typical Allan deviation of a 1:1 measurement of graphene against GaAs/AlGaAs heterostructure at $\nu=2$ for a measurement current of $100\ \rm\mu A$. (Adapted from Ref.~\cite{Janssen2012})}
\end{figure}

Figure~\ref{labelfig6} shows the Allan deviation for a 3.5~h measurement of graphene against GaAs at $\nu=2$ of the CCC bridge in Ref.~\cite{Janssen2012}. The Allan deviation is very convenient for a resistance ratio measurement involving periodic current reversals as it leads directly to the expected measurement resolution for a given measurement time ($\tau$). From the figure we see that the first data point for a $\sim 40$~s measurement gives an uncertainty of around 4 parts in $10^9$. The Allan deviation decreases as $1/\sqrt{\tau}$ expected for white noise. After 3.5~h measurement time a relative uncertainty of 2 parts in $10^{10}$ is achieved. There is still room for improvement here; if the two CCC's can be made to operate at their optimum noise performance, a 40~s measurements would give a relative uncertainty slightly better than 1 part in $10^{9}$ and 6 parts in $10^{11}$ after 3.5~h (blue dot and dashed line in fig.~\ref{labelfig6}). 

\subsection{Quantized Hall resistance in terms of the ohm}
The determination of the value of $h/e^2$ in terms of SI units is a rather complex experiment and is not routinely performed. The derived SI unit for ohm is $\rm m^2 kg s^{-3}A^{-2}$ and depends on four of the seven base units. In practice, the route to realise the ohm does not involve the ampere realization but rather relies on a powerful electrostatics theorem by Thompson and Lampard \cite{Thompson1956}. The theorem states that the change in cross-capacitance of a specific four-electrode device as a function of change in guard electrode only depends on frequency and length which both can be realized with very small uncertainty. From impedance we can get to resistance via a long and complex chain of a.c. bridges which to date has been done to a few parts in $10^8$ \cite{Bachmair2009}. 

The determination of $R_{\rm K}=h/e^2$ in SI units also allows one to determine the value of $\alpha$, the fine structure constant, because $\alpha={\mu_0ce^2}/{2h}$. In the SI, the permeability of vacuum $\mu_0$ and the speed of light $c$ are fixed quantities with $\mu_0=4\pi\times 10^{-7}\rm\ NA^{-2}$ and $c=299792458\rm\ ms^{-1}$. Indeed, the first report on the quantum Hall effect was presented as a new and more accurate determination of $\alpha$ \cite{vonKlitzing1980}. This determination of $\alpha$ can be compared with other determinations such as high precision measurements and calculations of the anomalous magnetic moment of the electron, gyromagnetic ratio of protons or the mass of neutrons which result in a combined value of the fine structure constant with an uncertainty of 3.2 parts in $10^{10}$~\cite{Mohr2008b}. 

Early on after the discovery of the quantum Hall effect it was realized that the stability and reproducibility of the QHE was much better than the ability to realize its value in SI units. Therefore in 1990 it was decided to assign a constant value to $R_{\rm K}$ called $R_{\rm K-90}$ with value $25812.807\ \Omega$ exactly without uncertainty \cite{Taylor1989}. This has allowed metrology laboratories to calibrate standard resistors with uncertainty typically in the range of 5 ppb or better \cite{Jeckelmann2001}. 

\subsection{Universality tests in conventional 2DEGs}
The fact that $R_{\rm K}=h/e^2$ means that the quantum Hall effect should be universal, by which we mean that the value of $R_{\rm K}$ is independent of the sample properties (carrier density, mobility, dimensions, Landau level index and host material of the 2DEG). However, there is no quantitative theoretical model which proves this and the problem has essentially been addressed experimentally. The method described in the previous section is limited in accuracy by the complexities of an SI realization~\cite{Mohr2008b} and a much higher precision can be achieved by directly comparing two quantum Hall devices against each other in a dimensionless null measurement.

The first universality test was reported by Hartland {\it et al.}~\cite{Hartland1991} who measured the ratio of the $\nu=2$ plateau in a GaAs/AlGaAs sample and the $\nu=4$ plateau in a Si-MOSFET sample and obtained $R_{\rm H}(\nu =2;{\rm GaAs})/R_{\rm H}(\nu =4; {\rm Si})=2[1-0.22(3.5)\times10^{-10}]$. Later, Jeckelmann {\it et al.}~\cite{Jeckelmann1997} improved on this result and reduced the uncertainty to $2.3\times 10^{-10}$. Subsequently, the universality has been tested by varying the mobility, channel width and plateau index number~\cite{Jeckelmann1995}. From the results one can conclude that in semiconductor systems the universality between different semiconductors has been established within a few parts in $10^{10}$.

Even better performance for the universality test could be achieved by using four quantum Hall devices in a Wheatstone bridge configuration developed by Schopfer and Poirier~\cite{Schopfer2007}. The QHR devices are connected in a series-parallel network using the multiple connection scheme developed by Delahaye and a CCC senses the imbalance in the bridge. A direct comparison of four GaAs devices has demonstrated an accuracy of  4 parts in $10^{11}$ and an uncertainty of a few part in $10^{12}$ is conceivable~\cite{Poirier2009}. One must note that here the quantum Hall devices are identical and well matched and so this experiment doesn't test the universality of the QHE between different systems. 

\section{Graphene, bandstructure, electronic properties and the quantum Hall effect}

Graphene is the first two-dimensional crystal available for experiments \cite{ Novoselov2005, Zhang2005, Novoselov2004,NovoselovPNAS2005}. When compared to quasi-2D systems fabricated at interfaces (e.g. Si/SiO$_2$ or AlGaAs/GaAs), graphene displays a peculiar linear E$(k)$ dispersion (instead of a parabolic) that modifies substantially the magnetotransport picture. The following sections present an overview of the electronic structure of graphene and its implications for magnetotransport.

\subsection{Crystal and electronic structure}

Graphene owes its two-dimensional nature to $sp^2$ hybridization of carbon atoms. The crystal backbone is formed by in-plane localized $\sigma$ bonds, between carbon atoms, and out-of-plane $\pi$ electrons delocalized over the entire crystal. Each carbon atom has three nearest neighbors, separated by $\sim 120^o$ giving graphene its characteristic honeycomb structure.

\begin{figure}
  \centering
 {\includegraphics[scale=.6]{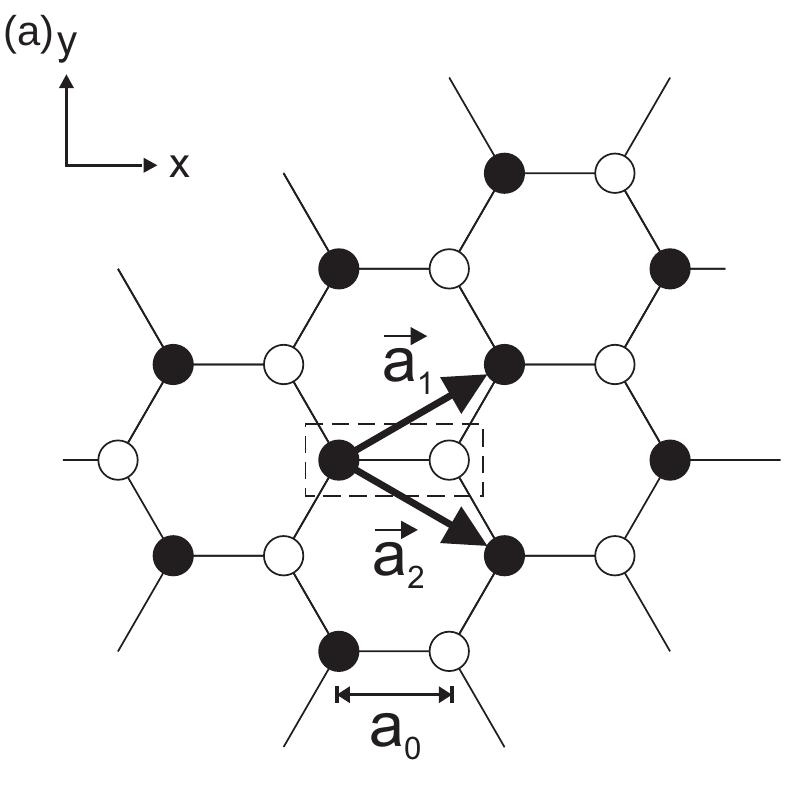}}
 {\includegraphics[scale=.6]{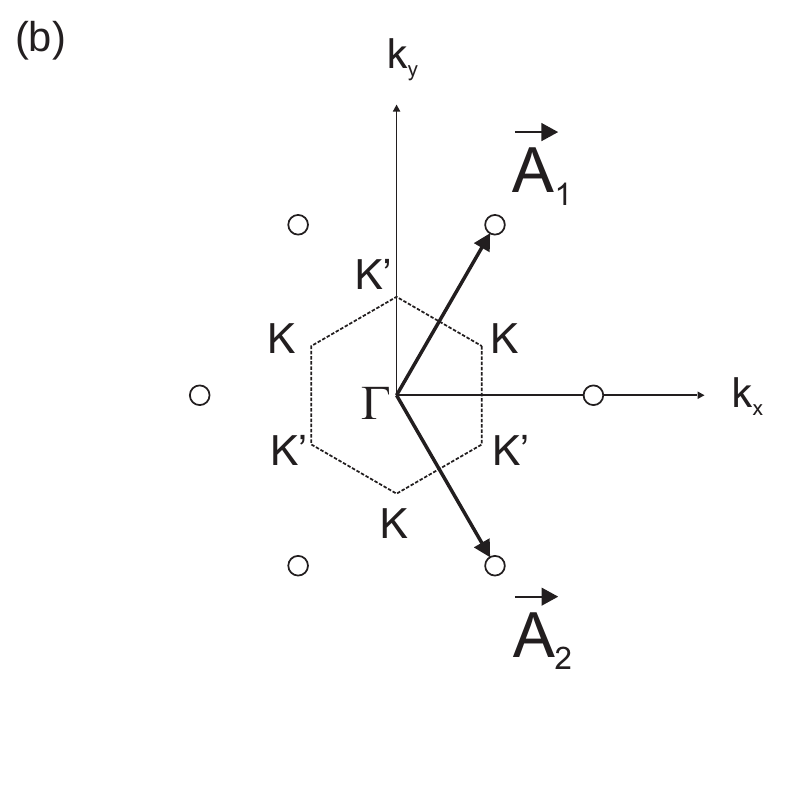}}
  {\includegraphics[scale=.6]{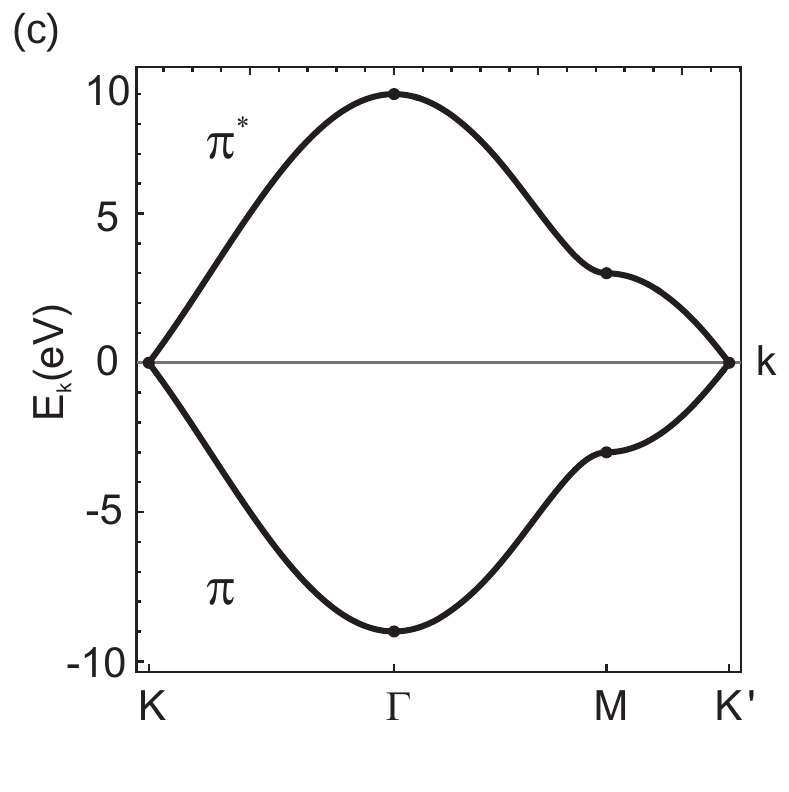}}
  \caption{(a) Graphene honeycomb structure (direct space); the unit cell contains two carbon atoms separated by $a_0=1.42\ \rm {\mbox{\normalfont\AA}}$. (b) Reciprocal lattice showing the first Brillouin zone (FBZ); six points at the corner of the FBZ fall into two groups of equivalent points, denoted as $K$ and $K$'. (c) Calculation of graphene bandstructure using a hoping parameter $t=-3$ eV; the bands are shown along the main crystallographic directions, including the $K$ and $K'$ points.}
  \label{labelfig7}
\end{figure}

The graphene lattice is described by a unit cell that comprises two atoms, $A$ and $B$, periodically arranged in a triangular lattice (fig. \ref{labelfig7}a). In real space, the primitive vectors are given by $\vec{a}_1 = \hat{x}a+\hat{y}b$ and $\vec{a}_2 = \hat{x}a-\hat{y}b $,  $a\equiv 3a_0/2$, $b \equiv \sqrt{3}a_0/2$ and the distance between nearest neighbors  $a_0=1.42$ {\AA}. The reciprocal lattice is constructed as $\vec{K}=M\vec{A}_1+N\vec{A}_2$, where $(M,N)$ are integers and the primitive vector in the reciprocal lattice $\vec{A}_1$ and $\vec{A}_2$ are determined from the condition $a_i\cdot A_j=2\pi \delta _{ij}$: $\vec{A}_1=\hat{x} (\pi/a)+\hat{y}(\pi/b)$ and $\vec{A}_2=\hat{x} (\pi/a)-\hat{y}(\pi/b)$ (fig. \ref{labelfig7}b).

Since graphene is a covalent solid, a good approximation of its electronic band structure can be found using a tight-binding description \cite{Wallace1947} with two basis functions per unit cell (one $p_z$ orbital per carbon atom). This is justified in materials with $sp^2$ bonding, in which the $\sigma$ bonds, localized in the graphene plane, are decoupled from the delocalized $p_z$ orbitals. Generalization from the unit cell to the entire solid is achieved by using Bloch function as ansatz. The band structure is described by:

\begin{equation}\label{labeleq7}
    E=\pm t \sqrt{1+4\cos(k_yb)\cos(k_xa)+4\cos^2(k_yb)}, 
\end{equation}
with $t$ is the hopping parameter.
The two bands (fig. \ref{labelfig7}c), a consequence of having two basis functions per unit cell, are symmetric about $E=0$. In neutral graphene the Fermi level lies at $E=0$, and all the states with $E<0$ are filled while those with $E>0$ are empty. In reciprocal space, the location of the charge-neutrality point ($E=0$) is found by setting Eq. \ref{labeleq7} equal to 0; with $k_xa=0$, the condition $E=0$ is satisfied at the six corners of the first Brillouin zone, shown in fig. \ref{labelfig7}(c). The 6 corners fall in two groups of three equivalent points, differing only by a reciprocal lattice vector. These two groups represent then two non-equivalent points, named for crystallographic convention $K$ and $K'$. 

Graphene is thus a zero band gap semi-metal, with the Fermi level located at the intersections between the valence and conduction ($\pi$ and $\pi^*$) bands, which are located at the $K$ and $K'$ points in $k$-space.

\subsection{Low-energy spectrum}

The results from the tight-binding model are particularly interesting close to the $K$ and $K'$ points in reciprocal space, where the energy dispersion $E\propto k$, in contrast to $E\propto k^2$ as in conventional electrical semiconductors. Around these $K$, $K'$ points, the energy dispersion of carriers is similar to that of ultra-relativistic particles with zero rest mass $m_0$,   $E(\vec{k})=\sqrt{m_0^2c^4+c^2\hbar^2k^2}= c\hbar k$; under these conditions the Schr\"odinger equation for Bloch electrons reduces to the 2D Dirac equation \cite{CastroNeto2009,Katsnelson2012}. The Dirac equation is thus used to describe the behaviour of carriers in graphene, which mimic massless Dirac Fermions around the $K$, $K'$ (\textit{Dirac}) points. 

By taking the $K$($K'$) point as reference and defining a vector momentum relative to this point as $q=k-K$ ($q'=k-K'$), the eigenfunctions in momentum representation for the pseudo-relativistic carriers at the $K$ and $K'$ points are
\begin{equation}
 \psi_{\pm,K}(q)=\frac{1}{\sqrt{2}} \left( e^{-i\theta_k/2} \pm e^{i\theta_k/2}\right)
\end{equation}
\begin{equation}
\psi_{\pm,K'}(q')=\frac{1}{\sqrt{2}}\left( e^{i\theta_k/2} \pm e^{-i\theta_k/2}\right)
\end{equation}
 with the $+/-$ signs corresponding to energies of the $\pi^*$ and $\pi$ bands. 
 
The wave functions at each Dirac point are thus described by a two component spinor, a linear combination of excitations arising from the A and B sublattice ($\pi$ and $\pi^*$ bands). The \textit{pseudo-spin}, quantified by the angle $\theta_k$, is in fact related to the vector momentum of the carriers and not to the real spin of electrons:

\begin{equation}\label{labeleq8}
\theta_k=\arctan\left(\frac{q_x}{q_y}\right)
\end{equation}
Since $\theta_k$ is a function of momentum, it follows that the wavefunctions at $K$ and $K'$ are related by time-reversal symmetry ($t \rightarrow -t$).

Another consequence of the pseudo-spin being associated to the vector momentum $q$,
is that when carriers move in a full circle 
the wavefunction changes sign (or equivalently, it accumulates an effective phase $e^{\pm i\pi}$). Taking as an example one of the components of $\psi_{+,K}(k)$, a change of $360^o = 2\pi$ results in $e^{-i(\theta_k+2\pi)/2}=e^{-i\pi}e^{-i\theta_k/2}=-e^{-i\theta_k/2}$. This is also called \textit{geometric} or \textit{Berry} phase, and it has interesting implications whenever electrons in graphene move along closed trajectories, as for cyclotron motion at high magnetic fields or self-crossing paths in quantum diffusive regime at low magnetic fields.

\subsection{Half-integer quantum Hall effect}

Novel and unique magnetotransport features are observed in graphene as a consequence of the facts that carriers: 1) can be modelled as having no mass, and 2) accumulate a Berry phase of $\pi$ when completing a full turn around the $K$, $K'$ points. To start with, the expression for the cyclotron frequency for Dirac Fermions is modified from that in conventional 2D systems \cite{Novoselov2005}:
\begin{equation}\label{labeleq9}
\omega_c=\sqrt{2}\frac{v_F}{\ell_B}=v_F\sqrt{\frac{2eB}{h}}
\end{equation}

Additionally, instead of using the Schr\"{o}dinger equation, the LL spectrum for chiral carriers in graphene is found by solving the Dirac equation in the presence of electric and magnetic field. The LL spectrum is given by \cite{Novoselov2005, Gusynin2005, McClure1957}:
\begin{equation}\label{labeleq10}
     E_{LL-Gr} = \pm \hbar \omega_c \sqrt{N} = v_F \sqrt{2\hbar eBN}
\end{equation}
with $N$ an integer number including zero and $v_F$ the Fermi velocity. The main differences with conventional 2D systems are:
\begin{itemize}
\item The energy spacing of LL in graphene depends on the magnetic field as $\Delta E_{LL} \propto\sqrt{B}$, instead of $\Delta E_{LL} \propto B$ as in conventional 2D systems.
\item Each LL in graphene can take twice as many electrons as LL do in conventional 2D systems. This four-fold degeneracy is due to spin-up/spin-down (as in conventional 2D systems), and valley degeneracy, $K$ and $K'$ (particular to monolayer graphene).
\item The special $LL$ in graphene at $E=0$ is shared equally by electrons and holes and. As a consequence it contains half as many states as the rest do.
\end{itemize}

\begin{figure}[th]
  \centering
  {\includegraphics[scale=.6]{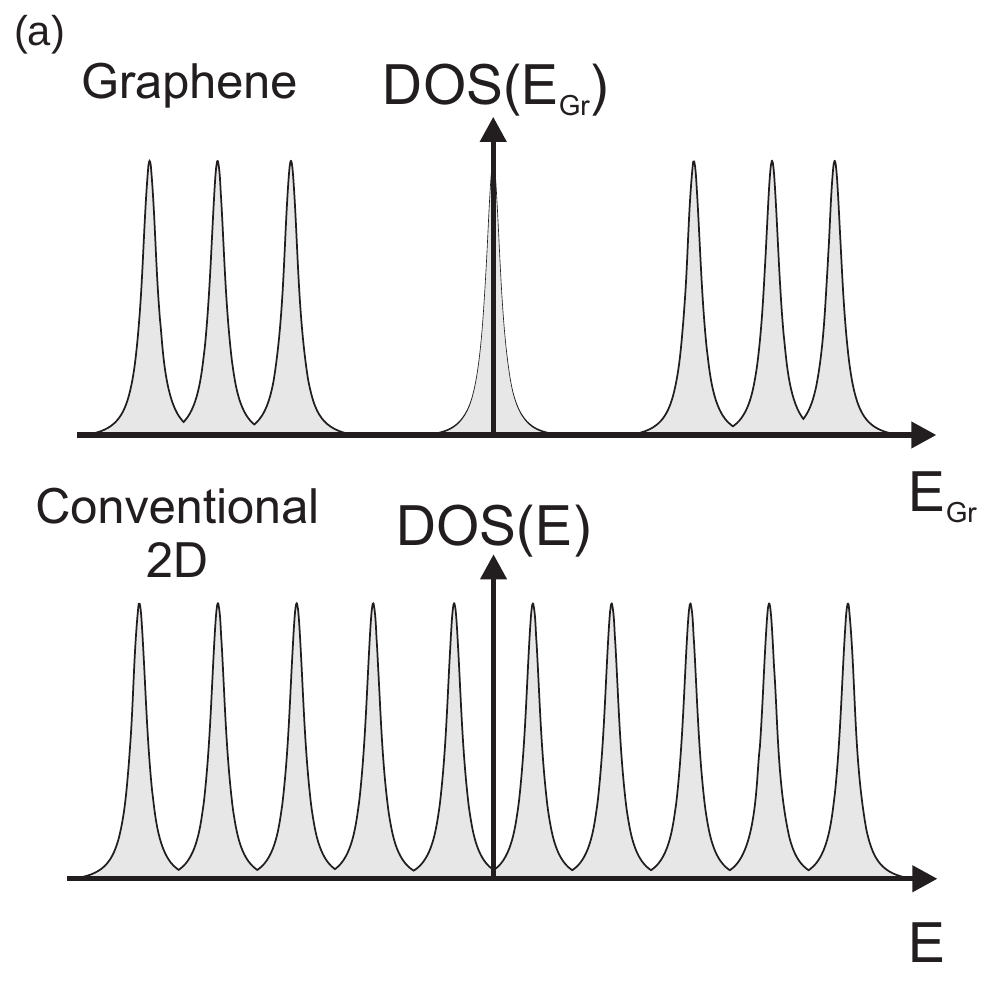}}
  {\includegraphics[scale=.85]{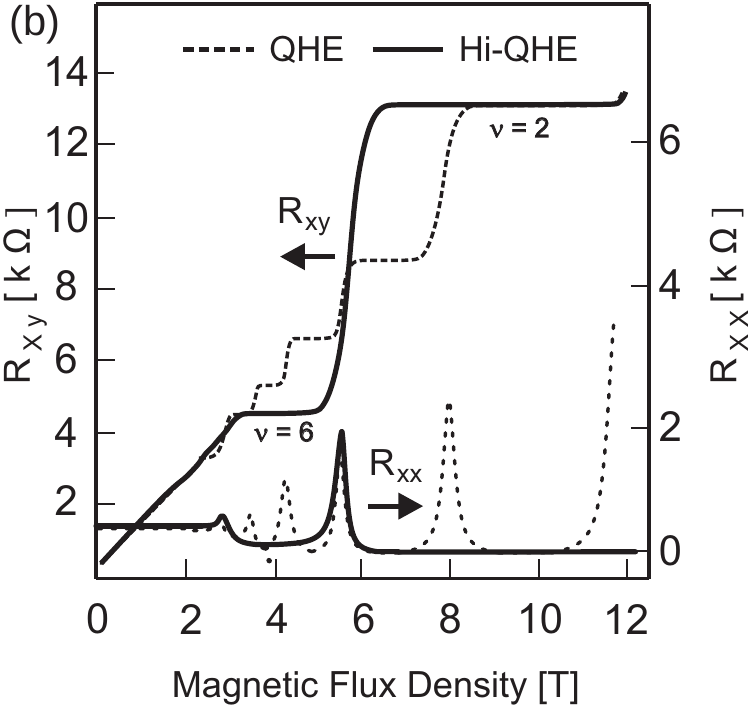}}
  \caption{(a) The sequence of Landau levels in graphene is unique; the energy spacing depends on the magnetic field as $\Delta E_{LL} \propto\sqrt{B}$, instead of $\Delta E_{LL} \propto B$ as in conventional 2D systems and there exists a LL at E=0, shared equally by electrons and holes. (b) Comparison of quantum Hall effect (QHE, dashed lines) in conventional 2D systems and the half-integer quantum Hall effect (Hi-QHE, solid lines). For graphene, only $\nu=2$ and $\nu=6$ are shown for clarity. }
  \label{labelfig8}
\end{figure}

This anomalous Landau level spectrum results in a peculiar sequence of plateaux in $\rho_{xy}$ in the quantum Hall regime, the so-called half-integer quantum Hall effect (fig. \ref{labelfig8}).  
\begin{equation}\label{labeleq11}
\rho_{xy}=\frac{h}{4e^2} (N+1/2)^{-1}= \frac{h}{e^2} (4N+2)^{-1}
\end{equation}

with $N$ an integer number including zero. 
This can be understood by recalling the case of conventional 2D systems, in which each (spin degenerate) Landau level can allocate $2eB/h$ electrons (per unit area) (Eq. \ref{labeleq5}). In addition to spin degeneracy, for graphene we need to take into account an additional double (valleys $K$ and $K'$) degeneracy. Thus, if we take into count only electrons (Fermi level $E_F>0$), the electron density $n_s$ corresponding to $N$ filled LL in graphene is:  
\begin{equation}\label{labeleq12}
n_s =  N\frac{2eB}{h}\vert_{K}+ N\frac{2eB}{h}\vert_{K'}+\frac{2eB}{h}\vert_{E=0}=\frac{4eB}{h}(N+1/2)
\end{equation}

By substituting the discretized density of states n$_s$ into the B-dependence of the transversal resistivity $\rho_{xy}$ in conventional Hall effect, we arrive at the sequence of plateaux for the half-integer quantum Hall effect:
\begin{equation}\label{labeleq13}
   \rho_{xy}=\frac{B}{en_s}=\frac{B}{e[4eB(N+1/2)/h]}=\frac{h}{4e^2(N+1/2)}
\end{equation}

The half-integer quantum Hall effect is the fingerprint of monolayer graphene, and can be used experimentally to prove that electronic transport occurs through a single graphene layer.

In bilayer graphene (two layers of graphene stacked on top of each other) the dispersion relationship becomes parabolic and the charge carriers behave like Dirac fermions with Berry phase $2\pi$~\cite{CastroNeto2009}. This leads to a QHE with resistance plateaux at

\begin{equation}\label{labeleq14}
   \rho_{xy}=\frac{h}{4e^2(N)}
\end{equation} 

where $N\geq 1$~\cite{McCann2006}. Thus the observation of a particular series of quantum Hall plateaux can be used to distinguish between mono- or bilayer graphene.

\section{Quantum Hall effect in exfoliated graphene}

\begin{figure}[th]
  \centering
  \includegraphics[scale=.3]{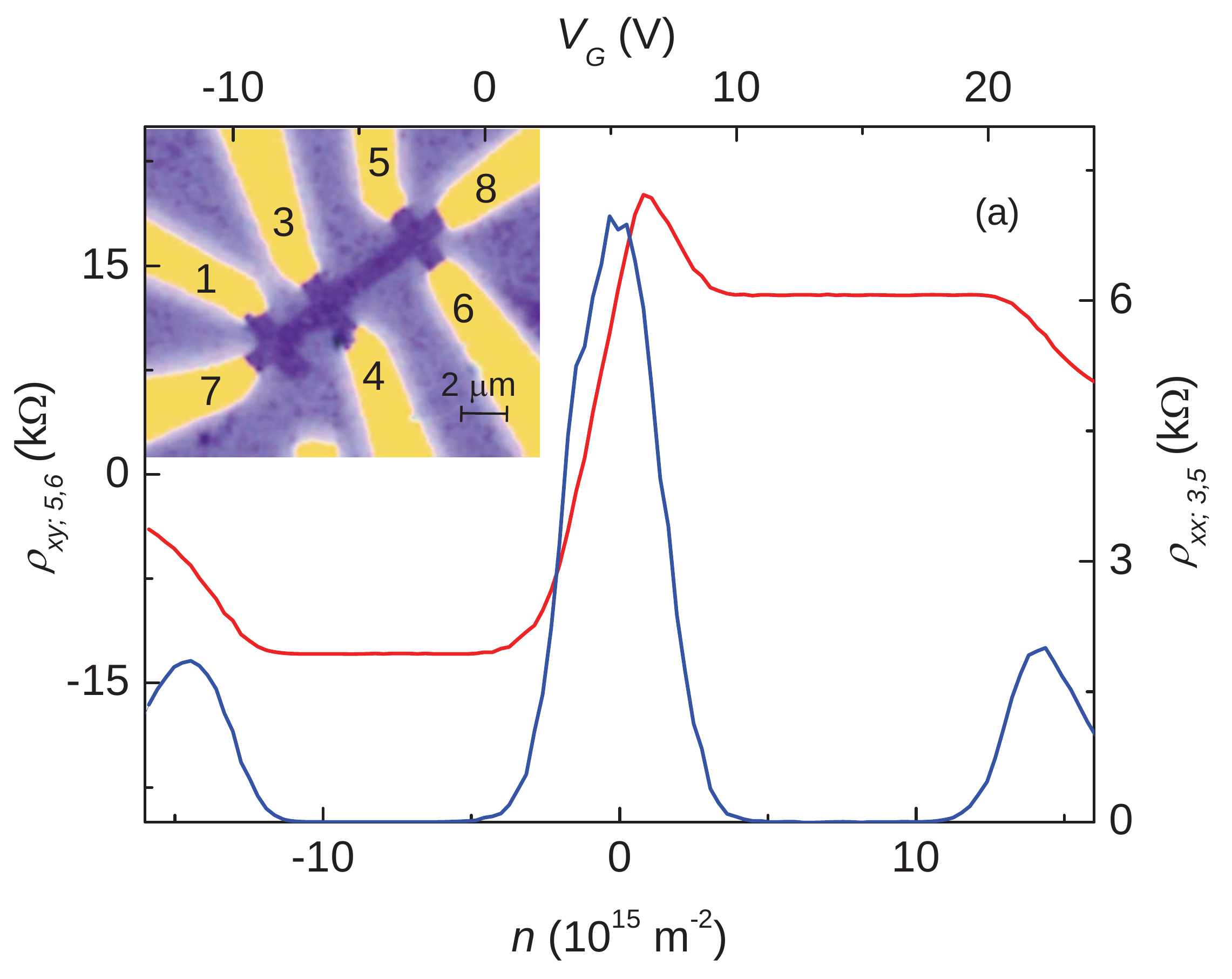}
  \includegraphics[scale=.3]{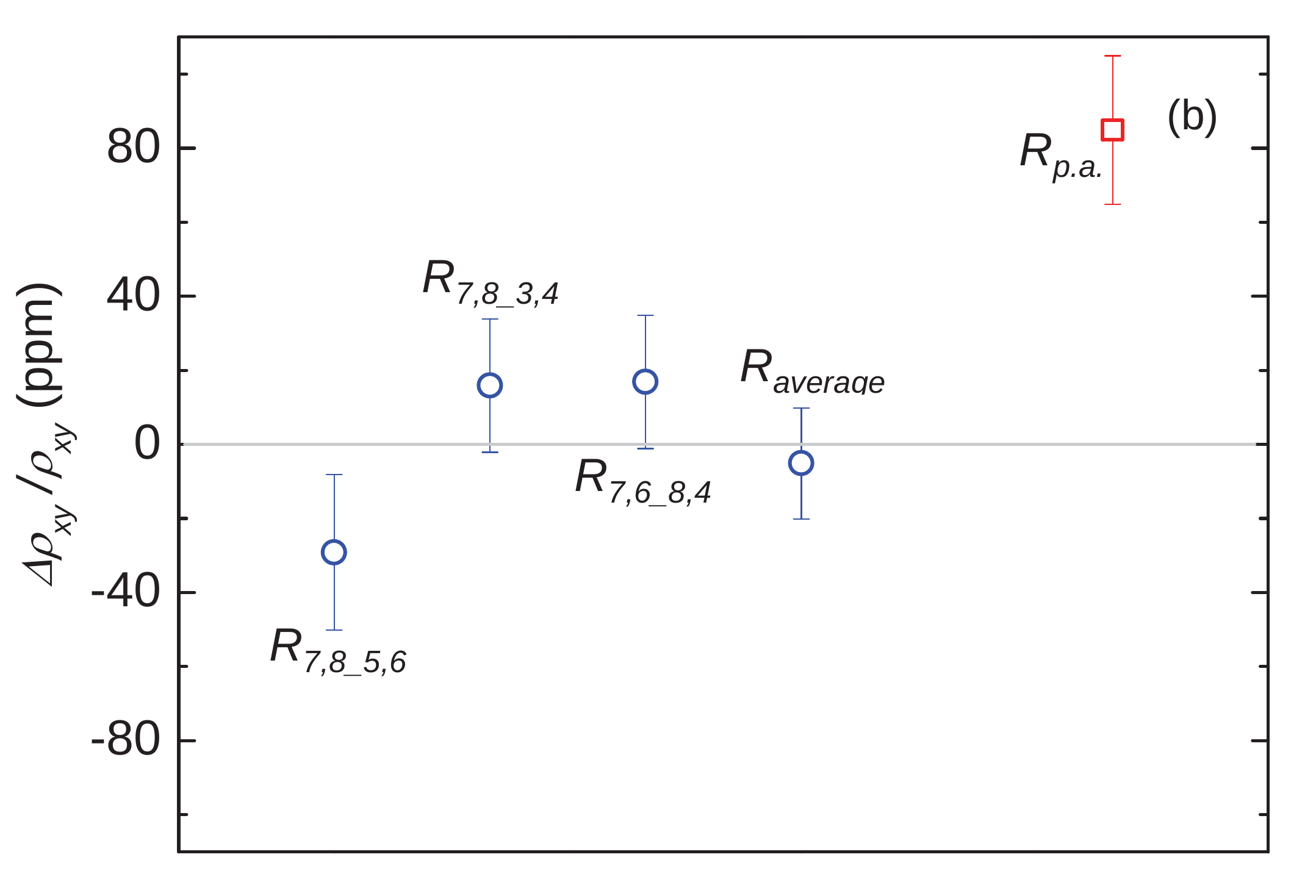}
  \caption{(Color online) The first accurate measurements of the quantized Hall resistance in exfoliated graphene. (a) Longitudinal resistivity $\rho_{xx}$ (blue) and Hall resistance $\rho_{xy}$ (blue) at $B-14\ \rm T$ and $T=0.35\ \rm K$ as a function of gate voltage. Inset: false color SEM of the device. (b) Deviations from quantization in ppm measured with a CCC and a bias current of $1.5\ \rm\mu A$ for different contact configurations. The red square is for a poorly annealed sample. (From Giesbers {\it et al.}~\cite{Giesbers2008})}\label{labelfig9}
\end{figure}

The accuracy of the resistance quantization in graphene was first tested by measuring its value in terms of a room temperature 100~$\Omega$ resistor with a CCC bridge described in the section~\ref{Requirements}. The 100~$\Omega$ resistor was measured prior and after the graphene measurements in terms of a standard well-characterised GaAs heterostructure device~\cite{Giesbers2008}. In such an indirect comparison the accuracy is ultimately limited by the short-term stability of the room-temperature resistor to a few parts in $10^9$. However, the first measurements by Giesbers {\it et al.} on a small exfoliated sample were at a much lower level of 15 parts in $10^6$ (see fig.~\ref{labelfig9}). This was attributed in part to the small breakdown current the small device could sustain (2.5~$\mu\rm A$ for a width of 1 $\mu \rm m$) which results in a small signal to noise ratio. One should note however that this small breakdown current still equates to a current density of 3.5~A/m which is equal or better than the best GaAs devices. A second limitation was the relatively high contact resistances (in the range of 1~k$\Omega$) which increases the noise in the measurements and leads to local heating~\cite{Jeckelmann2001}.

\begin{figure}
  \centering
  \includegraphics[scale=0.7]{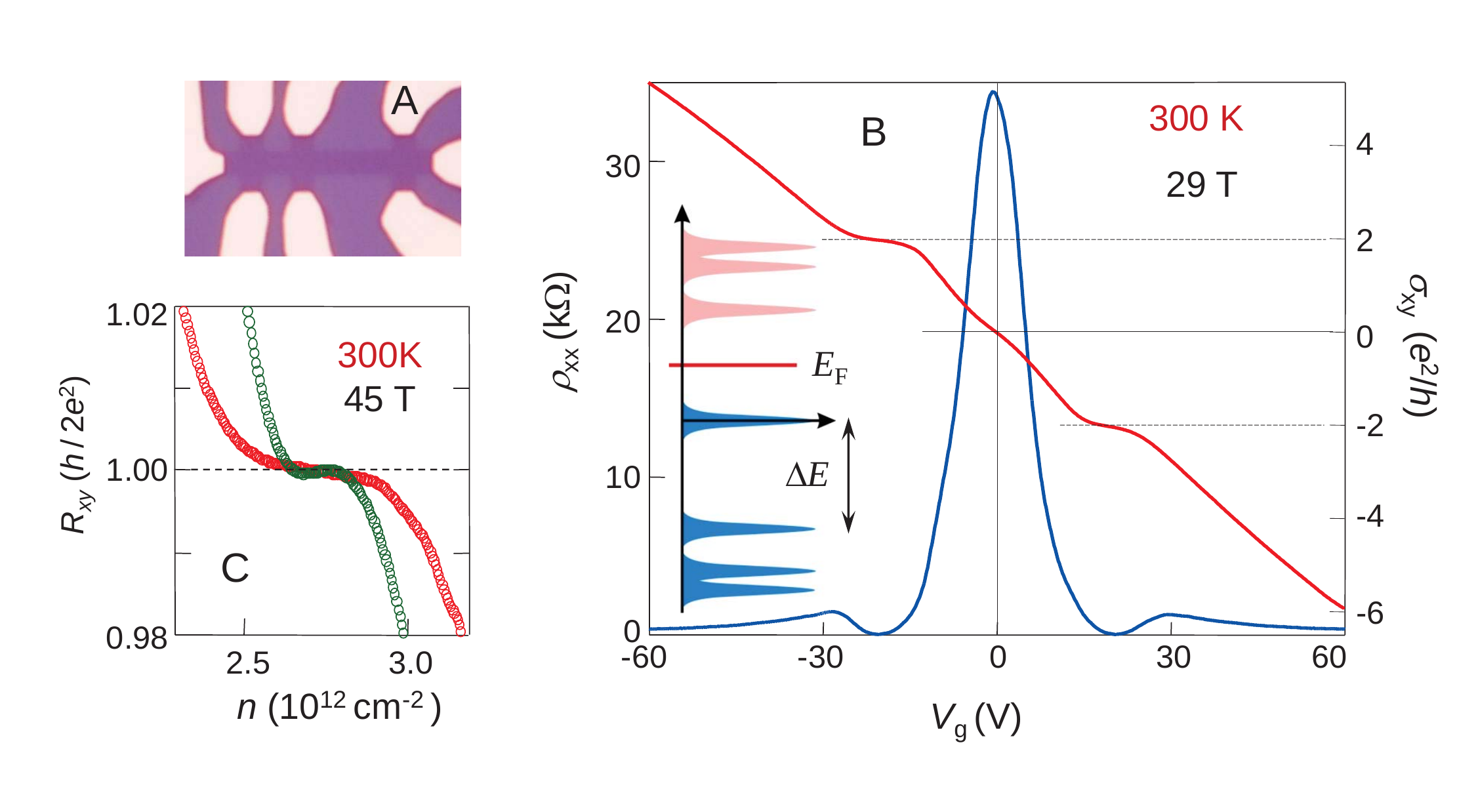}
  \caption{(Color online) Room temperature QHE in exfoliated graphene. (a) Optical micrograph of device. The width of the Hall bar is $2\ \rm\mu m$. (b) $\sigma_{xy}$ (red) and $\rho_{xx}$ (blue) as a function of gate voltage ($V_g$) in a magnetic field of $29\ \rm T$. (Inset) The LL quantization for Dirac fermions.(c) Hall resistance, $R_{xy}$, for electrons (red) and holes (green) at $45\ \rm T$. From Novoselov {\it et al.}~\cite{Novoselov2007}}\label{labelfig10}
\end{figure}

The unusual bandstructure of graphene results in a very large energy gap in the LL spectrum (see Eq.~\ref{labeleq10}) which implies that the QHE can be observed in graphene at much higher temperatures compared to traditional semiconductor systems. Room-temperature QHE was first demonstrated by Novoselov {\it et al.}~\cite{Novoselov2007} for an exfoliated graphene device in magnetic field in access of $20\ \rm T$ (see fig.~\ref{labelfig10}). The longitudinal resistivity approaches zero ($<10\ \Omega$) and the Hall resistance quantization was limited by the experimental accuracy of $\approx 0.2\%$ in these measurements. The mobility of the device was  $\approx 10000\ \rm cm^2V^{-1}s^{-1}$ which results in significant LL broadening and extremely large magnetic fields are required to observe the QHE at room-temperature. If the mobility could be improved, the magnetic field could be reduced and room-temperature QHE could be observed in conventional magnet systems.

\begin{figure}
\centering
	\includegraphics[scale=0.7]{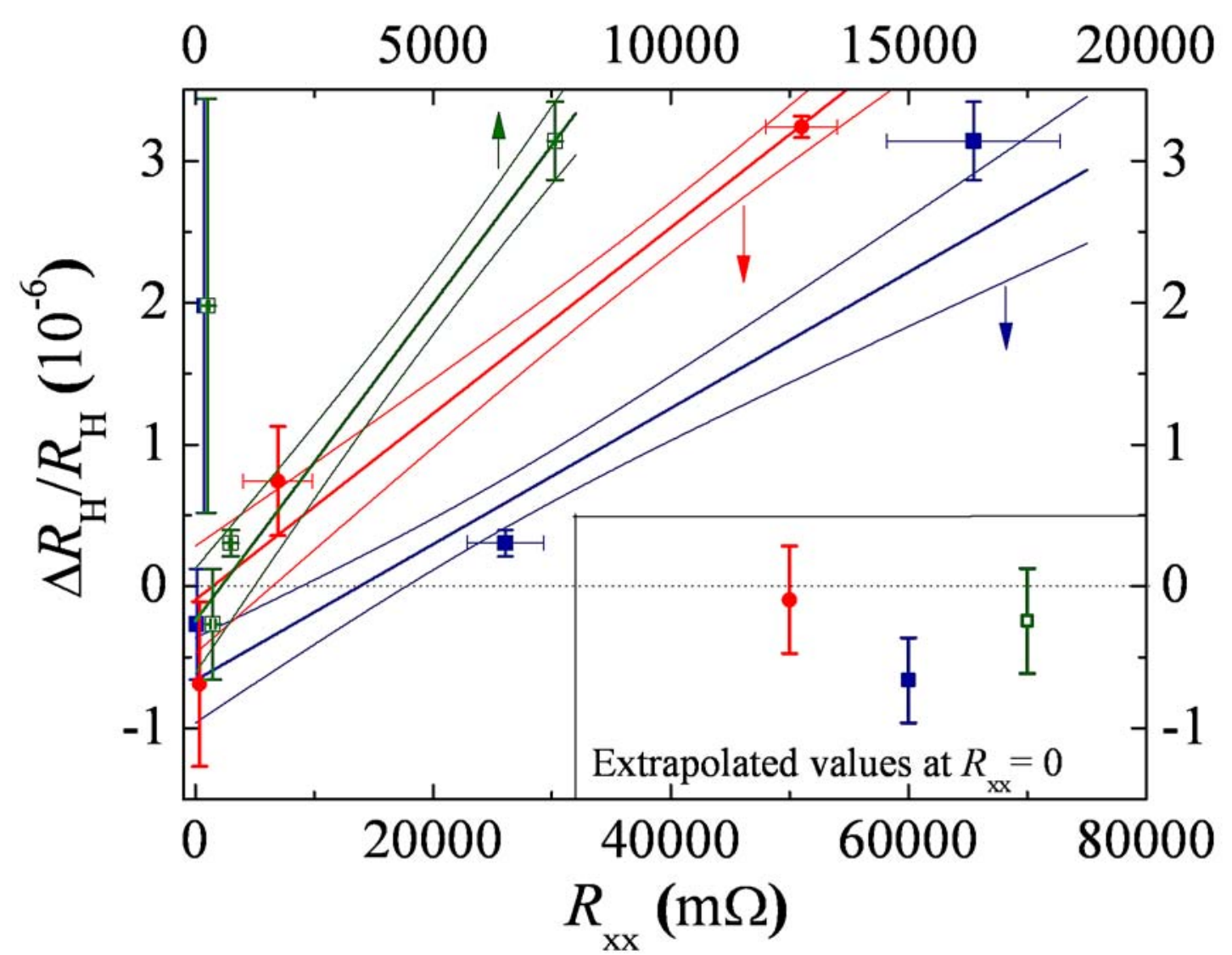}
	\caption{\label{labelfig11} (Color online) $\Delta R_{\rm H}/R_{\rm H}$ as a function of $R_{xx}$ for an exfoliated bi-layer graphene device. The different colors are data for different contact configurations. The extrapolated values of $\Delta R_{\rm H}/R_{\rm H}$ for $R_{xx}=0$ are shown in the inset. Errors bars correspond to measurement uncertainties given within one standard deviation, $1\sigma$.(From Guignard {\it et al.}~\cite{Guignard2012})}
\end{figure}

The group at Laboratoire Nationale de m\'{e}trologie et d'Essais (LNE) improved on the initial results by Giesbers by measuring both exfoliated monolayer and bilayer graphene samples with significantly lower contact resistances (between 10-500~$\Omega$)~\cite{Guignard2012}. They made detailed measurements of both the Hall and longitudinal resistances and used the empirical relation, $\Delta\rho_{xy}=k\rho_{xx}$ (with $k$ a constant), to extrapolate to $\rho_{xx}=0$, the dissipationless limit for which perfect quantization is expected (see the result for bilayer graphene in fig.~\ref{labelfig11}). Both mono and bilayer graphene devices were found to be quantized to within 5 parts in $10^7$. This accuracy is again limited by the small breakdown current that the exfoliated device could sustain (about 1~$\mu\rm A$ for a width of 2 $\mu \rm m$). The LNE group determined that their devices suffer from a large degree of charged impurity scattering which could limit the breakdown current and results in considerable charge density fluctuations.

\begin{figure}[th]
  \centering
  \includegraphics[scale=.7]{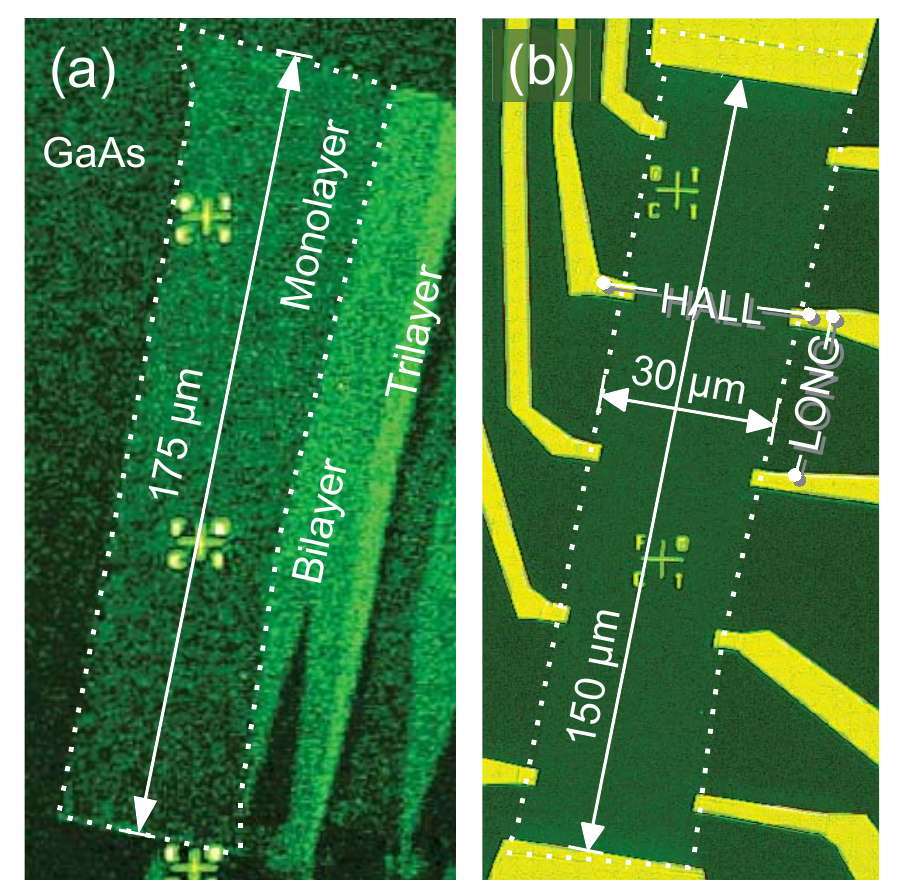}
  \includegraphics[scale=.9]{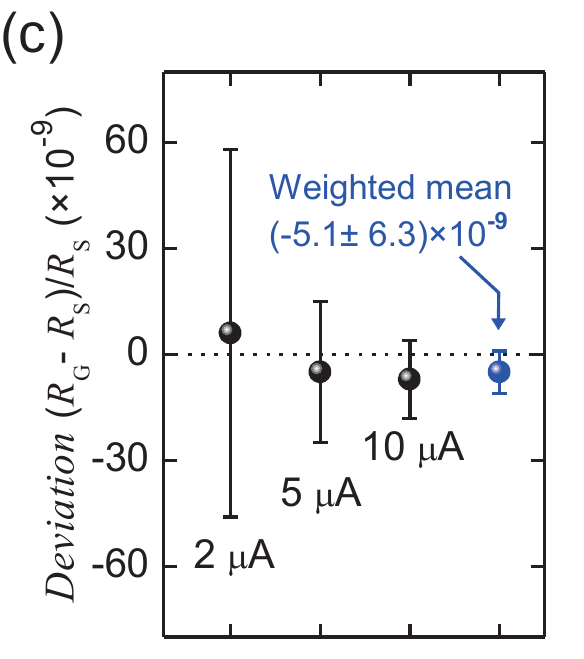}
  \caption{ (Color online) Quantum Hall resistance measurements on large area exfoliated graphene. (a) Large area unprocessed graphene flake on GaAs. (b) A complete device made from the monolayer marked by the dotted line. (c) Relative deviation between quantum Hall resistance and a standard resistor for different source-drain currents. The blue dot is the weighted mean of these deviations. (From Woszczyna {\it et al.}~\cite{Woszczyna2011}) }\label{labelfig12}
\end{figure}

Recently the group at the Physikalisch Technische Bundesanstalt (PTB) have produced very large (30~$\mu \rm m$ by 150~$\mu \rm m$) exfoliated graphene samples on top of a GaAs/AlAs heterostructure~\cite{Woszczyna2011} [see fig.~\ref{labelfig12}(a) and (b)]. GaAs was chosen as a substrate because its surface is much smoother than the more commonly used SiO$_2$ surface and its higher dielectric constant should improve the electrical screening of substrate defects. It was also found that the GaAs surface results in much larger exfoliated flakes which, they speculate, might be as a result of the stronger hydrophilic character of GaAs. The heterostructure also contained a strongly doped GaAs layer which serves as a back gate to control the carrier density. The device had a breakdown current of $\sim 15\ \rm \mu A$ (i.e. a breakdown current density of 0.5~A/m) and all contact resistances were below 10~$\Omega$. The reported uncertainty for their precision measurements~\cite{Woszczyna2012} was about 6 parts in $10^9$ [see fig.~\ref{labelfig12}(c)] for zero back gate voltage which is on a par with the early precision measurements in epitaxial graphene of similar size~\cite{Tzalenchuk2010}. The measurements were done at zero gate voltage because of a leakage between the graphene and back gate layer~\cite{Woszczyna2011}. This leakage implies that a parallel conduction path exists to the graphene layer and this could lead to a systematic deviation from $R_{\rm K}$ which would need to be eliminated in future.

It is very encouraging that accurate QHE results can be obtained with exfoliated graphene. However, the breakdown current and hence the achievable accuracy are at least an order of magnitude lower than those obtained in large area epitaxial graphene which is the topic of the next section.

\section{Graphene engineering for resistance metrology}

\subsection{Growth and Characteristics of as-grown epitaxial graphene}\label{CharGraphene}

The highest quality graphene is produced by the well-known mechanical exfoliation from graphite (more commonly known as sticky-tape method)~\cite{Novoselov2004}. Although perfect for scientific research, their small size limits the applicability in high-precision metrology because of the need for high current densities and small contact resistances. Large-area graphene can be produced by two distinct methods, chemical vapour deposition (CVD) and high-temperature sublimation of silicon carbide. 

With the CVD technique very large areas of good quality graphene can be produced by decomposing and graphitizing organic material on a hot metal surface. The metal acts as a catalyst for the reaction and typically copper is used as a substrate and methane as a carbon-containing gas. After growth, for most applications, the graphene needs to be transfered to a non-conducting substrate, a process which degrades the quality of the graphene by introducing doping and defects. Nevertheless, the Purdue group have demonstrated the QHE in a 7~mm by 7~mm graphene layer grown by CVD on copper, and then transferred to a SiO$_2$/Si substrate~\cite{Shen2011}. Low fabrication costs are one of the main advantages of the CVD technique, while its challenges are the reliable control of the nucleation sites and the development of reliable, scalable and non-destructive graphene transfer methods. Since the whole field is driven by the need to replace indium-tin-oxide in touch-screen displays by highly conductive graphene~\cite{Bae2010}, these challenges must be met soon. However, in view of metrological applications, the low cost of graphene samples is of less importance, since it is negligible in comparison with the measurement equipment. 

Growth of graphene on SiC occurs when the crystal is annealed at high temperatures ($\leq 1300\ ^\circ$C): the silicon sublimes and the carbon-rich surface re-crystalises to form graphene~\cite{Berger2004}. The quality and number of layers of graphene, critically depends on the precise growth conditions, crystal type, face and orientation. Compared to CVD, the electronic properties of the as-grown large area monolayer graphene produced by this method are often superior in terms of mobility and doping. 
The most commonly used polytypes of SiC are 4H-SiC and 6H-SiC, where H stands for hexagonal and the number refers to the number of basel planes (Si-C bilayers) in the unit cell. Both 4H-SiC and 6H-SiC crystals have two different faces, either terminated by carbon or silicon atoms. The reaction kinetics of silicon desorption is different for the carbon and silicon faces, it is much faster on the C-face compared to the Si-face.

Growth on the C-face leads to many graphene layers stacked on top of each other. These graphene layers are rotated with respect to each other and as a result are electrically decoupled to some extent and are therefore different from ordinary graphite~\cite{Hass2008}. Very high mobilities (of order $100000\ \rm cm^2V^{-1}s^{-1}$) have been achieved in multilayer graphene but the QHE has never been observed in this system~\cite{Darancet2008}. Recently the Georgia Tech group succeeded in growing a single monolayer graphene on the C-face of 4H hexagonal SiC~\cite{Wu2009} with mobilities in the range of $\rm 5000-20000\ \rm cm^2V^{-1}s^{-1}$ and hole carrier density in the range of $\rm (1-2)\times 10^{12}\ cm^{-2}$. A clear half integer QHE effect was observed in this system, although the metrologically relevant parameters were not explored in this work. 

Growth on the Si-face is much slower and allows better control of the thickness through the growth parameters; fabrication of monolayer graphene is possible on this face. A number of groups have succeeded almost simultaneously in observing the QHE by growing monolayer graphene on the Si-face in a high vacuum~\cite{Shen2009} and in an argon atmosphere~\cite{Tzalenchuk2010,Jobst2010,Tanabe2010}. Typically mobilities in the range $\rm 500-5000\ \rm cm^2V^{-1}s^{-1}$ are obtained and electron carrier densities in the range $\rm (0.5-10)\times 10^{12}\ cm^{-2}$. Jobst {\it et al.}~\cite{Jobst2010} have made a careful study of the spread in device parameters ($\mu$ and $n_s$) by analyzing data from over 50 devices. They found that for their growth temperature of around 1600$^\circ$C and the 6H-polytype, the average mobility is $900\ \rm cm^2V^{-1}s^{-1}$ and the carrier density is $1\times 10^{13}\ \rm cm^{-2}$. This high carrier density makes it impossible to use these devices for quantum Hall effect resistance metrology directly (the $\nu=2$ magnetic field would be larger than 100~tesla). 

The SiC crystals are nominally cut at a 0$^{\circ}$ angle but in reality a slight miss-cut will lead to a number of atomically sharp step edges separated by flat terraces across the crystal surface. The Erlangen group investigated the effect of these step edges by placing small Hall bars either entirely on a terrace or deliberately across a number of step-edges. Surprisingly, there was no noticeable effect on the mobility~\cite{Jobst2010}; theoretical analysis of the role of steps on SiC on the charge carrier scattering in graphene, together with the review of experimental data is given in~\cite{Low2012}. Growth parameters and degree of miss-cut determine the substrate morphology and the geometry of bilayer islands, which are nucleated along the substrate steps \cite{Emtsev2009} and affect mobility. 

Virojanadara {\it et al.}~\cite{Virojanadara2008} found that the use of an inert gas during growth allows for an increase in the annealing temperature ($\sim$2000$^\circ$C) by reducing the sublimation rate of silicon carbide. The advantage of the higher growth temperature is that it leads to atomically uniform graphene with larger graphene domains ($\sim 50\ \rm \mu m$). Tzalenchuk {\it et al.}~\cite{Tzalenchuk2010} investigated the properties of large-area epitaxial graphene devices grown on the 4H-polytype using this technique.  Hall bar devices of different sizes, from 160~$\rm \mu m \times 35\ \rm \mu m$ down to 11.6~$\rm \mu m \times 2\ \rm \mu m$ were produced, 20 on each 0.5~$\rm cm^2$ wafer, using standard electron beam lithography and oxygen plasma etching (Figure~\ref{labelfig1}). Atomic force microscopy (AFM) images revealed that the graphene layer covers the substrate steps like a carpet, preserving its structural integrity. Contacts to graphene were produced by straightforward deposition of 3~nm of Ti and 100~nm of Au through a lithographically defined mask followed by lift-off, with the typical area of graphene-metal interface of
$\rm 10^4\ \mu m^2$ for each contact. The manufactured material was n-doped, with the measured electron concentration in the range of $\rm (5-8)\times 10^{11}\ cm^{-2}$, mobility about $\rm 2400\ cm^2V^{-1}s^{-1}$ at room temperature and between 4000 and $\rm 7500\ cm^2V^{-1}s^{-1}$ at 4.2~K, almost independent of device dimensions and orientation with respect to the substrate terraces similar to results obtained in Ref.~\cite{Jobst2010}. 

\begin{figure}
\centering
	\includegraphics{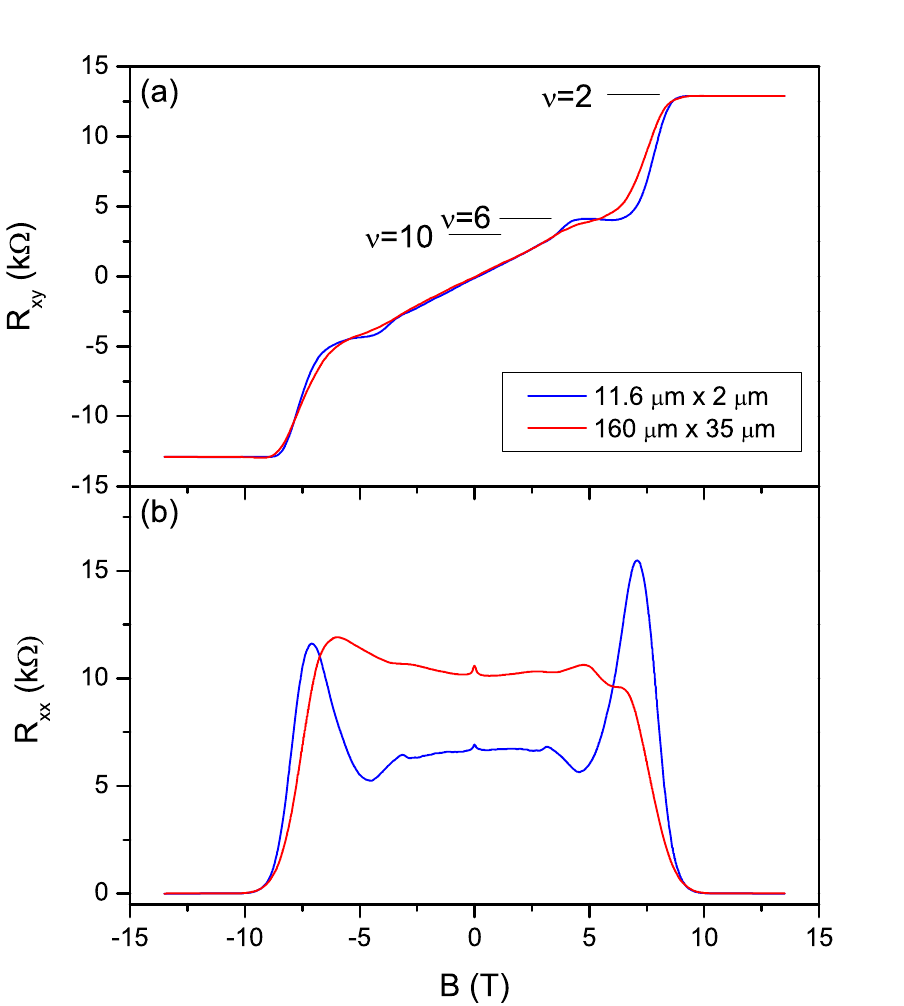}
	\caption{\label{labelfig13} (Color online) Transverse (a) and longitudinal (b) resistance of a small and large graphene on SiC device at $T=4.2\ \rm K$ with $1\ \rm \mu A$ current. (Adapted from Ref.~\cite{Tzalenchuk2010})}
\end{figure}

Figure~\ref{labelfig13} shows the longitudinal (dissipative) resistance, $R_{xx}$ and transverse (Hall) resistance, $R_{xy}$ of a $2\ \rm\mu m$ wide Hall bar at $4.2\ \rm K$ and magnetic field up to $14\ \rm T$. At high magnetic field one can clearly identify two QHE plateaux, at $R^{(0)}_{xy}=R_{\rm K}/2\ (N=0)$ and $R^{(1)}_{xy}=R_{\rm K}/6\ (N=1)$ corresponding to the filling factors $\nu=2$ and $\nu=6$, respectively. In graphene $\nu=2$ corresponds to the fully occupied zero-energy Landau level $(N=0)$ characterised by the largest separation $v_F\sqrt{2\hbar eB}$ (where $v_F\approx10^8\ \rm cms^{-1}$ is the Fermi velocity in graphene) from other Landau levels in the spectrum and hence the Hall resistance quantization is particularly robust. This plateau appears in the field range 9-12~T, depending on the carrier concentration (which was not controlled during these experiments) and is accompanied by a vanishing (within the noise of the measurement system) $R_{xx}$. The $N=1$ plateau, at $\nu=6$, is not so flat, and $R_{xx}$ develops only a weak minimum. There is also a trace of a structure corresponding to $\nu=10$. The observed sequence of Hall plateaux confirms that the material studied is indeed monolayer graphene. At low magnetic fields Shubnikov-de Haas oscillations can be observed, as well as a weak localization peak characteristic of the phase coherence of electrons in a disordered system (these effects have been analyzed comprehensively in  Ref.~\cite{Lara-Avila2011b}).

The magneto-transport measurements on a much larger, $160\ \rm\mu m\times 35\ \mu m$ Hall bar device are also shown in fig.~\ref{labelfig13}. A substantial positive magnetoresistance at low fields, which was absent in the smaller sample, indicated that the carrier concentration varied along the larger sample. Because of this, the $\nu=6$ feature in $R_{xx}$ in the larger sample was less prominent. Nevertheless, despite the inhomogeneity of the carrier density, the Hall resistance plateau at $R^{(0)}_{xy}=R_{\rm K}/2\ (N=0)$ was accompanied by vanishing longitudinal resistance $R_{xx}$. Importantly, the large-area device had a low contact resistance, $R_c\approx 1.5\ \Omega$, to the graphene layer and, as compared to smaller devices, could sustain a much higher current before QHE breaks down. Since larger breakdown current affords higher precision measurements in the QHE regime, the $160\ \rm\mu m\times 35\ \mu m$ Hall bar device was chosen for metrological measurements at a magnetic field of $14\ \rm T$ (This field was the maximum available and is still far from the $B = 17.5\ \rm T$ where the filling factor would be exactly $\nu= 2$ for this sample).

\begin{figure}
\centering
	\includegraphics{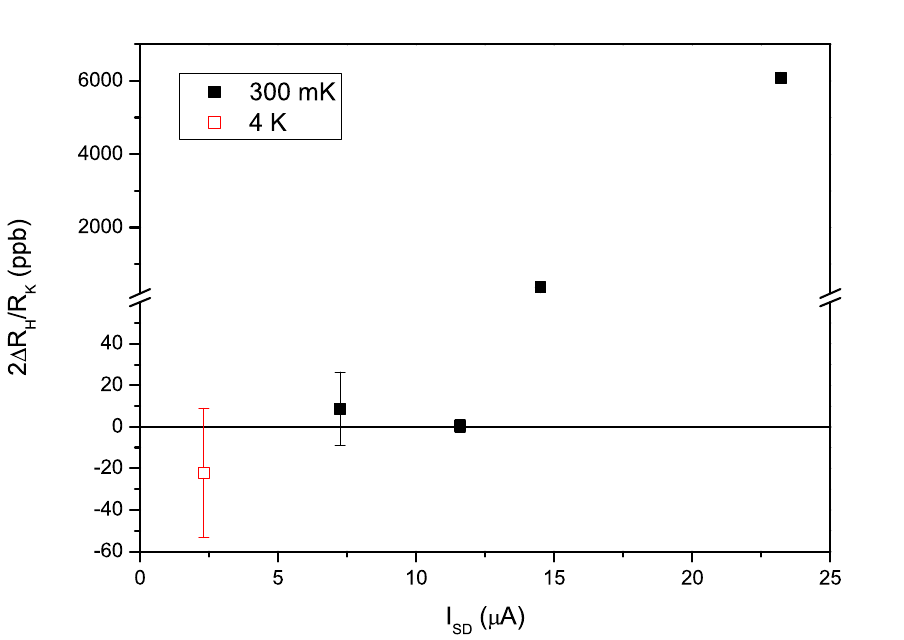}
	\caption{\label{labelfig14} (Color online) Precision measurement for a large graphene on SiC sample. The accuracy is expressed as the ratio between the measured quantum Hall resistance $R_{\rm H}$ and the value of $R_{\rm K}/2$ in parts in $10^9$ (ppb). The accuracy is dominated by the data point at $11.6\ \rm\mu A$ collected over 11 hours. The data points for higher currents are above the breakdown current where the device is no longer quantised. (Adapted from Ref.~\cite{Tzalenchuk2010})}
\end{figure}

The accuracy of the Hall resistance quantization in graphene was established in measurements traceable to the GaAs quantum Hall resistance standard using a calibrated $100\ \Omega$ resistor (see fig.~\ref{labelfig14})~\cite{Tzalenchuk2010}. The optimal conditions at 300~mK were obtained for a source-drain current of $11.6\ \rm\mu A$, 15\% below the breakdown current established in the measurements of $R_{xx}$. The quantization accuracy $+0.4\pm 3$ parts in $10^9$ inferred from the measurements was a four orders of magnitude improvement on the previous best result in exfoliated graphene \cite{Giesbers2008}. Graphene was still accurately quantized at 4.2~K, however, at this temperature the measurement current had to be reduced to $2.3\ \rm\mu A$, which increased the uncertainty of the data accumulated over a comparable time interval.  The precision measurement presented in Ref.~\cite{Tzalenchuk2010} readily puts epitaxial graphene quantum Hall devices in the same league as their semiconductor counterparts. Note that this accuracy was obtained on a sample, although large by graphene standards, substantially smaller than the semiconductor devices used for calibration and without optimization. Ideally, precision measurements are made at exactly integer filling factor because the breakdown current peaks at this value~\cite{Jeckelmann2001}. It is clear from the results presented in fig.~\ref{labelfig13} that this condition is not met for practical magnetic fields below 15~T. Further progress demands the ability to engineer the carrier density well below the $\rm 1\times 10^{12}\ cm^{-2}$-level and control the uniformity of charge distribution in epitaxial graphene (discussed in next section). 

Importantly, all twenty Hall bar devices fabricated on the chip shown in fig.~\ref{labelfig1}(a) demonstrated the quantum Hall effect, despite the fact that the steps on the substrate crisscrossed the devices. This tells us not only that the graphene is continuous on these steps, but also that the bilayer patches, which are often nucleating along the substrate steps, did not cross even the narrowest $1\ \rm \mu m$ Hall bar. Bilayer patches crossing the sample cause scattering of the edge states and destroy the resistance quantization~\cite{Lofwander2012,Schumann2012}. Recent AFM studies show wide variations of the bilayer patch sizes and shapes, starting from submicrometer islands to a few micrometer long stripes~\cite{Emtsev2009,Burnett2012}. While the growth of single-layer graphene on the Si side of SiC is under control in many laboratories, the control of bilayer morphology is a more elusive task, and more work is required here.

\subsection{Carrier density control in epitaxial graphene}

The work on graphene/SiC system described above has indicated a serious opportunity to use graphene in metrological applications. To reach the ideal $\nu = n_sh/(eB) = 2$ plateau, corresponding to the topologically protected $N=0$ Landau level at a feasible magnetic field, a low carrier density $n_s$ in graphene is required. Therefore control of the carrier density in graphene is of utmost importance for metrology.

Carrier density control in exfoliated graphene, transferred onto a SiO$_2$/Si substrate, is relatively easy. By applying a gate voltage to the doped silicon back gate, the carrier density can be changed by several orders of magnitude from electron to hole carriers right through the Dirac point~\cite{Novoselov2004}. A similar technique can be employed for graphene grown by CVD after transfer onto an oxidized silicon wafer~\cite{Shen2011}. For epitaxial graphene grown on a thick highly insulating substrate this method is not possible and other top and bottom gating techniques have been developed to control the charge carrier density.

As mentioned in the previous section, epitaxial graphene grown on Si-face of SiC is always strongly n-type doped with carrier densities typically in the range of $10^{12}-10^{13}\ \rm cm^{-2}$. The doping of the graphene is caused by the so-called ``dead layer'' or ``buffer layer'' of carbon atoms in between the SiC substrate and graphene~\cite{vanBommel1975,Riedl2007,Varchon2007,Mattausch2007,Emtsev2008,Qi2010}. This layer is non-conducting and characterized by a $6\sqrt{3}\times 6\sqrt{3}$ supercell of the reconstructed surface of sublimated SiC. Missing or substituted carbon atoms in various positions of such a huge supercell in the dead layer create localized surface states with a broad distribution of energies within the bandgap of SiC~\cite{Riedl2010}. Kopylov {\it et al.}~\cite{Kopylov2010} developed a theoretical model of the charge transfer in epitaxial graphene to estimate the as-grown carrier density and degree of variation that can be induced by a top gate. 

\begin{figure}
\centering
\includegraphics[scale=0.6]{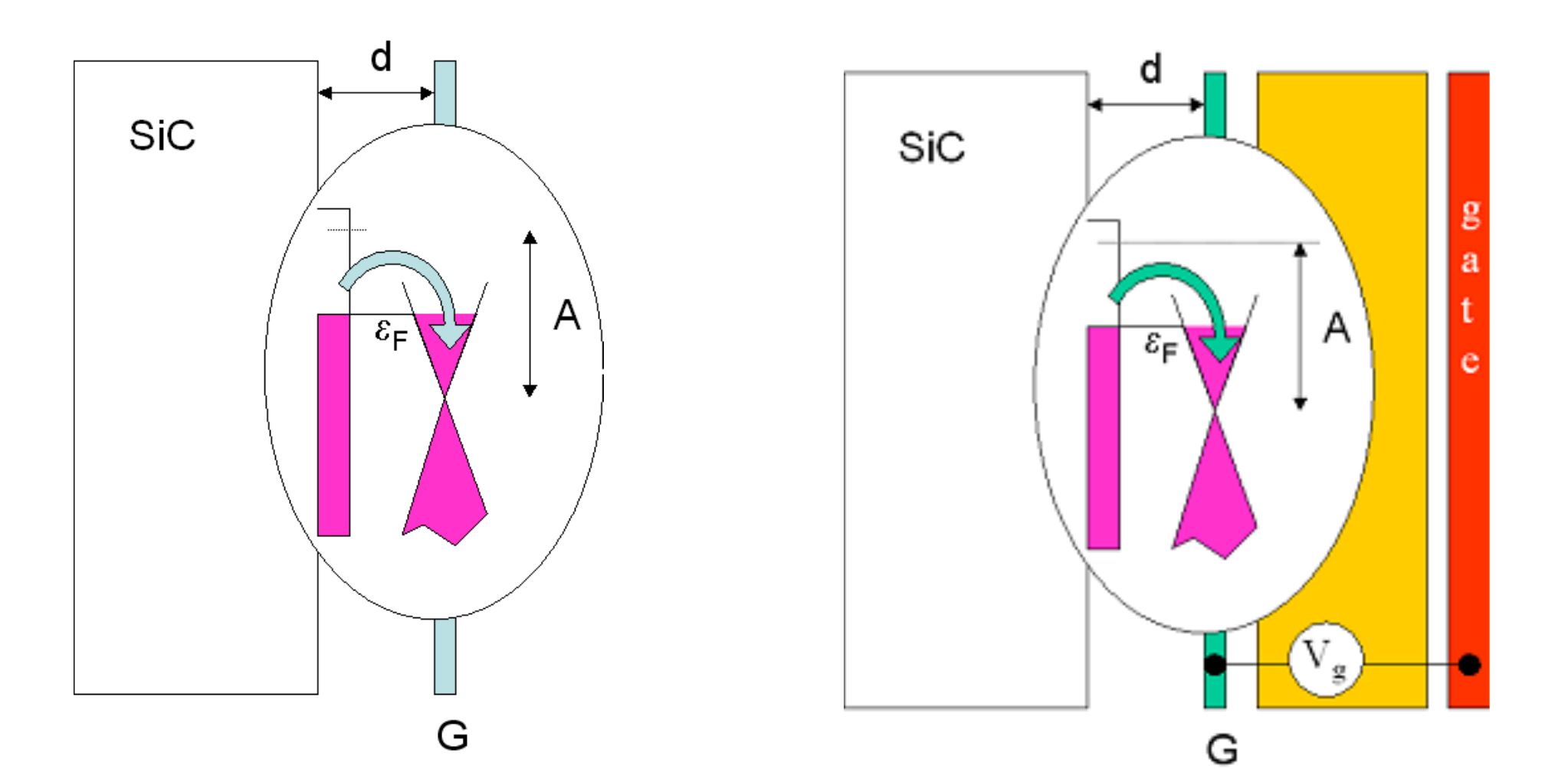}
\caption{\label{labelfig15} (Color online) Band structure model for charge transfer between SiC and bare/gated graphene. (Adapted from Ref.~\cite{Tzalenchuk2011})}
\end{figure}

The charge transfer results from the work function difference, $A$, between graphene and the combination of buffer layer and SiC substrate (fig.~\ref{labelfig15}). Assigning a density of states $\gamma$ to the donors in the buffer layer and $\rho l$ to the donors in the bulk ($\rho$ is volume density of donors and $l$ the depletion length), the charge balance equation can be written as,

\begin{eqnarray}\label{labeleq15}
\gamma [A-\frac{de^2n_s}{\varepsilon_r\varepsilon_0}-\epsilon_{F}]+\rho l=n_s. \nonumber\\
\end{eqnarray}

Here, $\epsilon_{F}=v_F\hbar\sqrt{\pi n_s}$, is the Fermi energy, and $\varepsilon_r\varepsilon_0/d$ is the geometrical capacitance per unit area of the graphene layer (here $\varepsilon_r$ and $\varepsilon_0$ are the relative permittivity and dielectric constant respectively and $d$ is the distance between the graphene and buffer layer). Typical values are $A=0.40\ \rm eV$, $d=0.3\ \rm nm$ and $\gamma= 5 \times 10^{12}\ \rm eV^{-1}cm^{-2}$. For an electrostatically gated device the charge is distributed between graphene and the gate ($n_g$), so we need to substitute $n_s$ by $(n_s+n_g)$ in Eq.~\ref{labeleq15}. $n_g=CV_g/e$ is determined by the gate-to-graphene capacitance per unit area, $C$, and voltage, $V_g$.
Solving this equation~\cite{Kopylov2010} results in $n_s\approx 10^{13}\ \rm cm^{-2}$ for zero gate voltage, which is close to the typical density of graphene doping observed in many recent studies of epitaxial graphene~\cite{Emtsev2009,Jobst2010,Coletti2010}.

The above consideration suggests that the reduction of donors on or just under the SiC surface is crucial for the use of epitaxial graphene on SiC in gated devices. We note that graphene growth at 2000$^\circ$C~\cite{Tzalenchuk2010} (a substantially higher temperature than in other published reports) reliably produced densities of $n_s\sim 10^{12}\ \rm cm^{-2}$ much lower than $10^{13}\ \rm cm^{-2}$. Tzalenchuk {et al.}~\cite{Tzalenchuk2011} speculate that high-temperature annealing of SiC compensates the surface donor states, possibly through surface segregation of impurities (e.g. B or N) abundant in bulk SiC crystals, or merely reduction of the number of surface defects.

\begin{figure}
\centering
\includegraphics[scale=0.2]{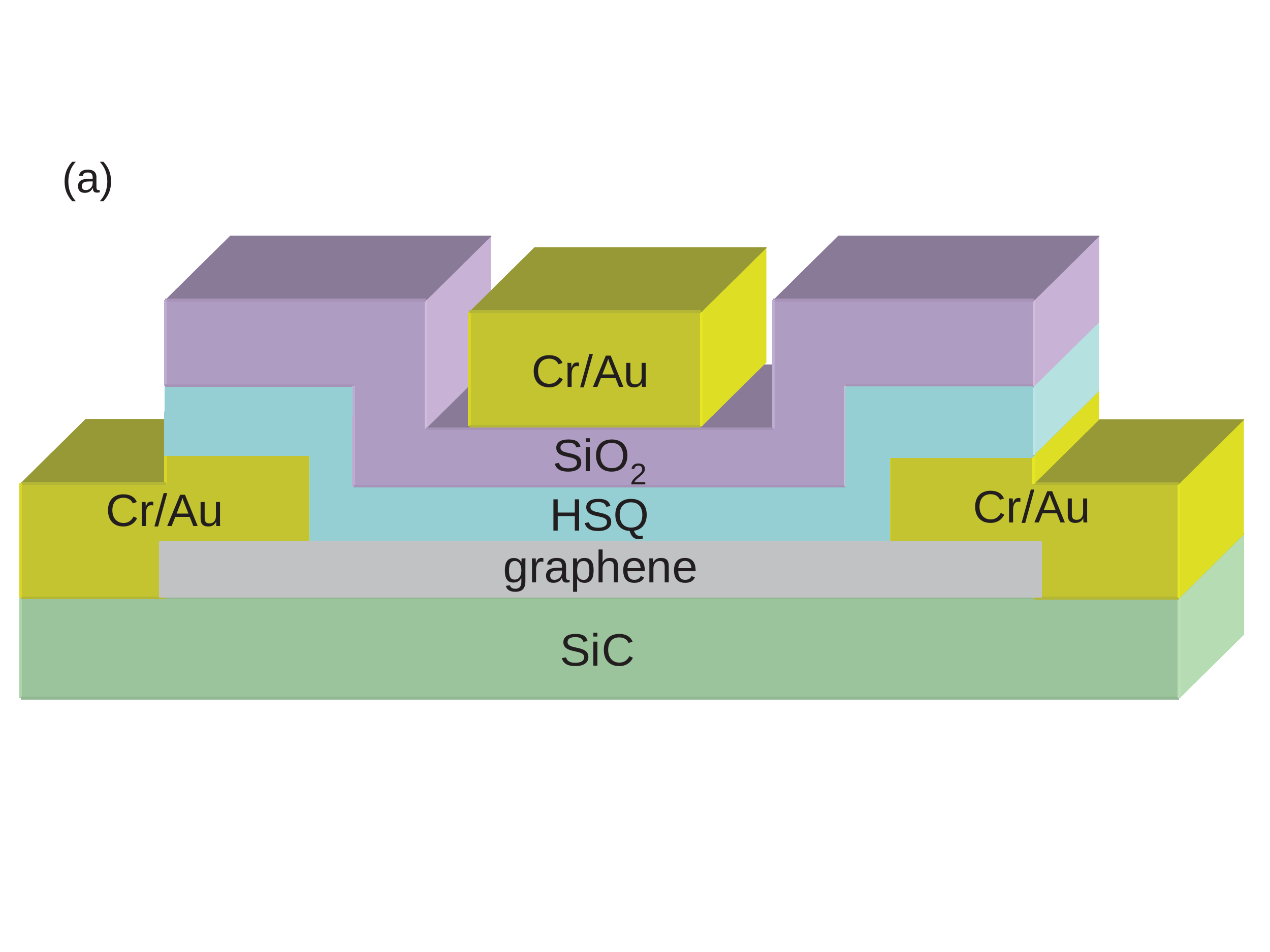}
\includegraphics[scale=0.2]{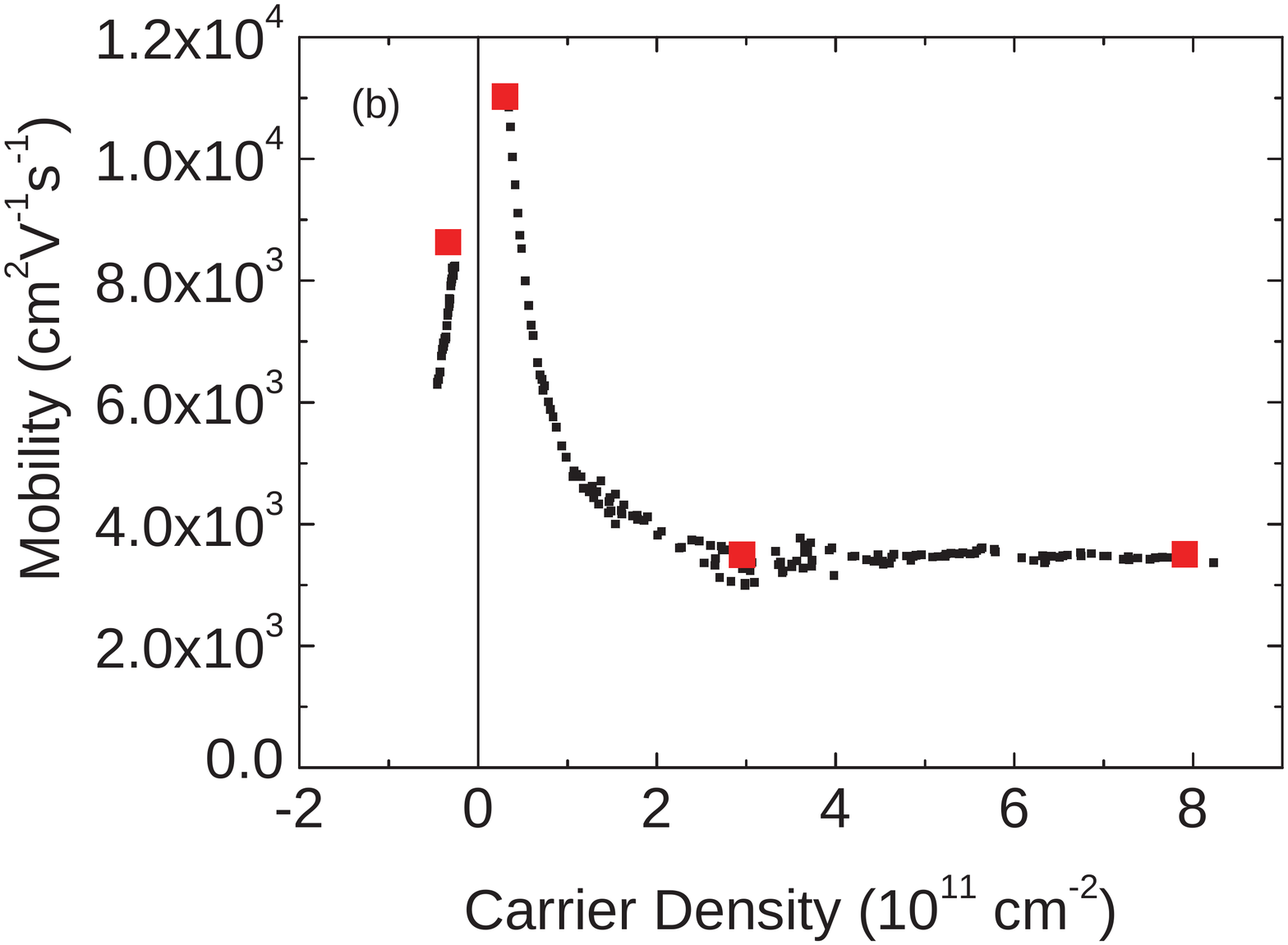}
\includegraphics[scale=0.2]{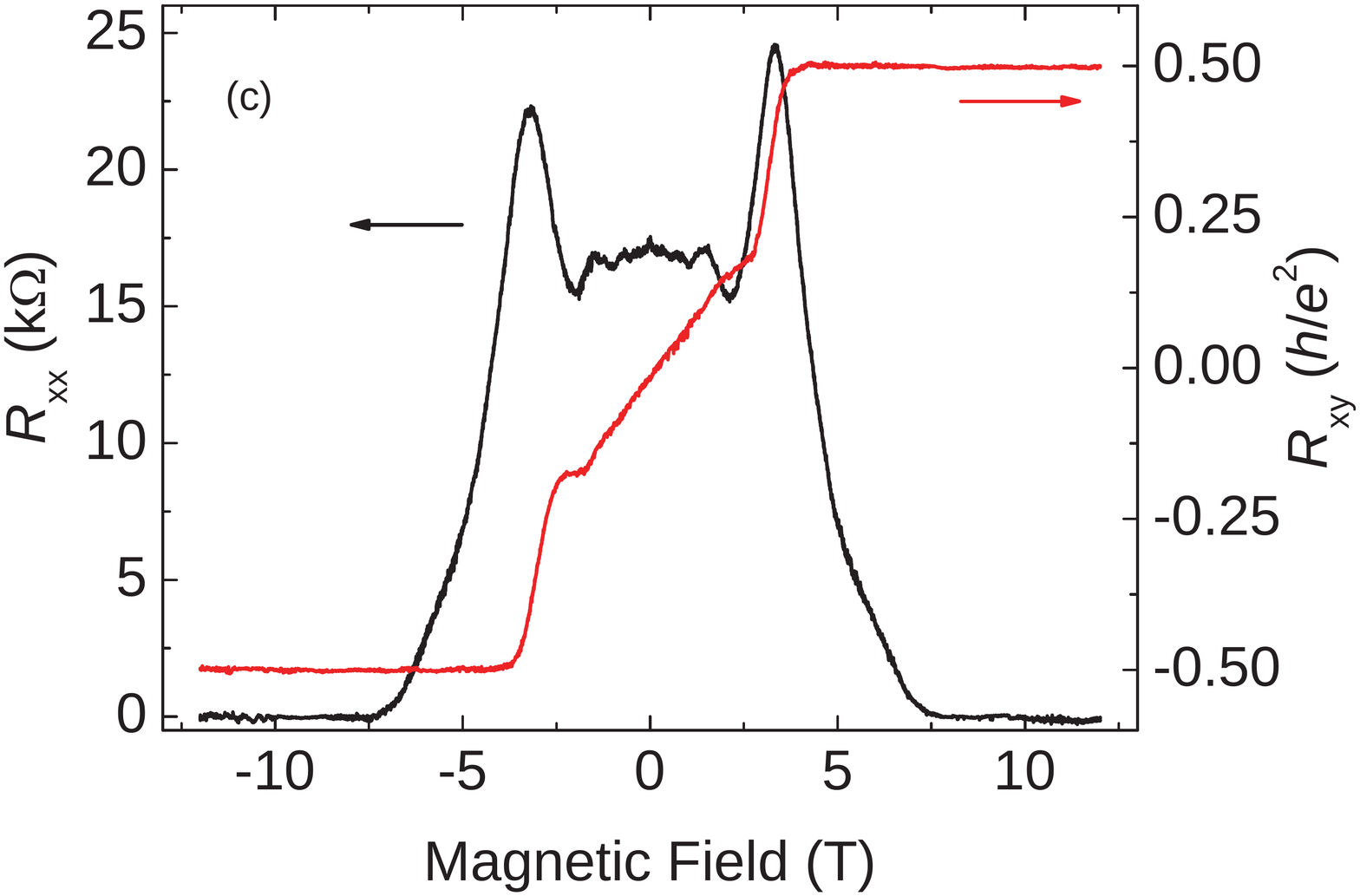}
\caption{\label{labelfig16} (color online) Top-gated graphene on SiC device. (a) Schematic cross-sectional view. (b) Mobility as a function of carrier density. For the black data points the carrier density is derived from $R_{xy}$ at $\pm0.5\ \rm T$ and the carrier mobility from $R_{xx}$ at $B=0\ \rm T$. For the red data points the carrier density is derived from the slope of $R_{xy}$ between -0.5 to 0.5~T. (c) $R_{xx}$ and $R_{xy}$ as a function of magnetic field at a gate voltage of -15~V where the carrier mobility and the carrier concentration were $3500\ \rm cm^2V^{-1}s^{-1}$ and $3\times 10^{11}\ \rm cm^{-2}$, respectively. (From Tanabe {\it et al.}~\cite{Tanabe2010})}
\end{figure}

Top gate technology has been demonstrated for epitaxial graphene by a number of groups. Tanabe {\it et al.}~\cite{Tanabe2010} produced a top gate stack of hydrogen silsequioxane (HSQ) and SiO$_2$ on top of a graphene Hall bar [see fig.~\ref{labelfig16}(a)]. The HSQ layer acted as an insulating layer and also protected the graphene from damage during processing. The metallic gate was formed by evaporating a layer of Cr/Au on top of the SiO$_2$. The mobility was unaffected by this processing and the carrier density could be changed from $10^{12}\ \rm cm^{-2}$ electrons through the Dirac point to $10^{11}\ \rm cm^{-2}$ holes [see fig.~\ref{labelfig16}(b)]. A clear half-integer QHE could be observed in these devices as demonstrated in figure~\ref{labelfig16}(c) Shen {\it et al.}~\cite{Shen2009} used an 30~nm Al$_2$O$_3$ gate dielectric deposited via atomic layer deposition on a 1~nm oxidized Al seeding layer, to create a gate stack. In this device a modest 50\% change in electron carrier density could be achieved. 

\begin{figure}
\centering
\includegraphics[scale=1.2]{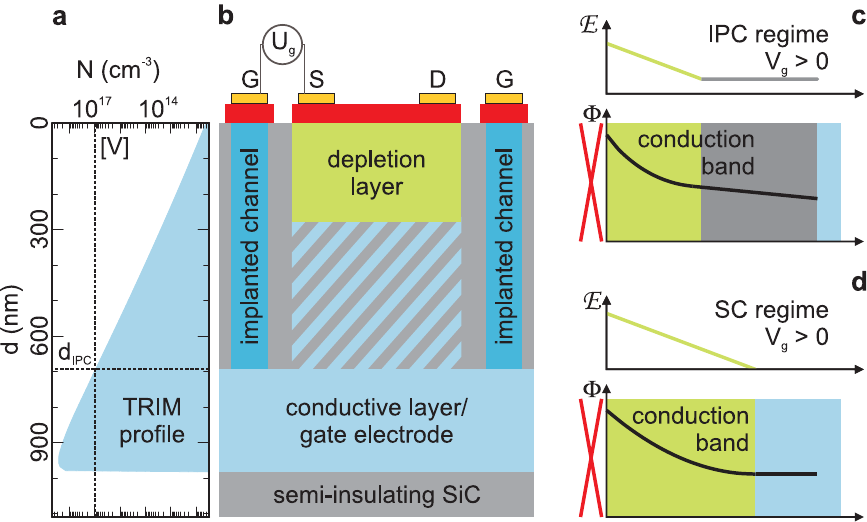}
\caption{\label{labelfig17} (Color online) Schematic of the Erlangen bottom gate device. A conductive layer is fabricated at 700~nm below the surface via implantation of nitrogen ions. (a) Simulation of the implantation profile. (b) Schematic of the bottom gate device with source (S), drain (D) and gate (G)electrodes on the graphene (red). The electronic properties of the shaded area depent on temperature and implantation dose. For low temperatures and low implantation dose the shaded area acts as an insulating layer and the gate behaves as an implanted plate capacitor (IPC). Band profile shown in (c). For high dose and high temperatures the gate behaves as a Schottky contact with band profile shown in (d).  (From Waldmann {\it et al.}~\cite{Weber2011}))}
\end{figure}

Back gate technology on Si-face epitaxial graphene devices was first demonstrated by the Erlangen group~\cite{Weber2011}. In this method a conducting layer is created some 700~nm below the surface of the SiC crystal through implantation of nitrogen ions prior to graphene growth (see fig.~\ref{labelfig17}). At low implantation dose this layer acts as a back gate and the carrier density can be changed through the Dirac point if the graphene layer is decoupled from the buffer layer through intercalation of hydrogen.  Without the intercallation step the donor states in the buffer layer will pin the Fermi level and effectively screen the graphene layer making the back gate ineffective. Also at low temperatures the conductivity of the implanted layer is strongly reduced and again the gate becomes ineffective. At high implantation dose or high temperatures the gate is much more effective and the carrier density can be changed over a wide range through the Dirac point.
Recently, using the nitrogen implantation technique, Jounault {\it et al.}~\cite{Jouault2012} demonstrated the QHE for both electrons and holes in a bottom-gated epitaxial graphene sample on the C-face of SiC. Because the graphene layers in C-face SiC are almost decoupled there is no need for the hydrogenation step.

A disadvantage of the gating technique is that an external voltage source has to be permanently connected to the device to maintain a constant carrier density. A solution would be a static gate in analogy to semiconductor programmable nonvolatile memory devices. These devices are essentially transistors with one extra floating, isolated gate sandwiched between the control gate and the semiconductor channel. Charge can be transferred to the floating gate by an electric pulse on the control gate and stored there isolated almost indefinitely, until intentionally leaked through the dielectric, e.g. activated by UV light. In other implementations of the nonvolatile memory devices UV light is used for writing in which case thermal activation can be used for erasing.

\begin{figure}
\centering
\includegraphics[scale=0.8]{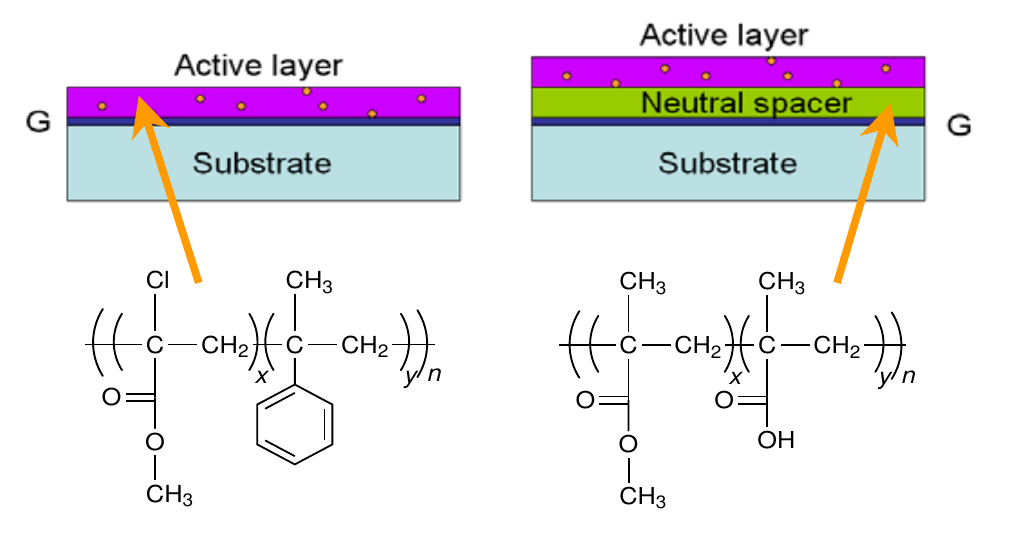}
\includegraphics[scale=0.8]{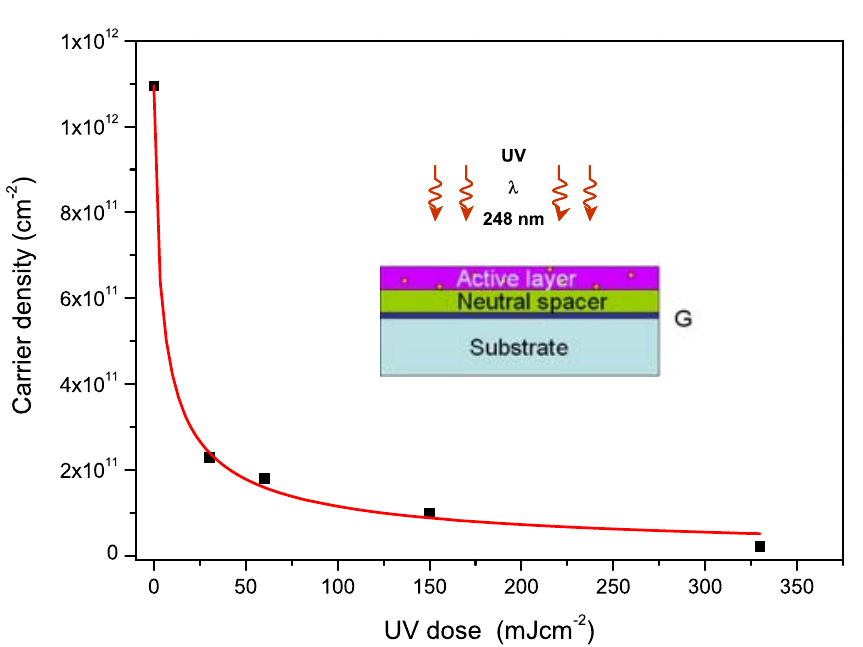}
\caption{\label{labelfig18} (Color online) Chemical gating of graphene on SiC. In direct chemical gating electrons in graphene can scatter on the ions in the adjacent active layer (left). When graphene is encapsulated in a polymer bilayer the active polymer (ZEP520A) is separated from graphene by a neutral spacer (middle). Carrier density as a function of UV exposure dose (right). (From Lara-Avila {\it et al.}~Ref.~\cite{Lara-Avila2011})}
\end{figure}

Lara-Avila {\it et al.}~\cite{Lara-Avila2011} demonstrated nonvolatile control of the carrier density by placing a polymer heterostructure of PMMA/MMA followed by another layer of ZEP520A, chosen for its ability to provide potent acceptors under deep UV light, on top of an epitaxial graphene device (both are commonly used clean-room photolithography resists). The layout of the heterostructure and the chemical formulae of the polymers are shown in fig.~\ref{labelfig18}(a); the ZEP520A is the floating gate which is isolated from the graphene by the PMMA/MMA layer which acts as the spacer. In a sample with the initial carrier density $n_s\approx 1.1\times 10^{12}\ \rm cm^{-2}$, subsequent exposures to UV at 248~nm wavelength up to the dose $330\ \rm mJcm^{-2}$ decreased the low-temperature electron density by 50 times down to $2\times 10^{10}\ \rm cm^{-2}$ [fig.~\ref{labelfig18}(b)]. This resulted in a fivefold increase in carrier mobility up to $16000\ \rm cm^2V^{-1}s^{-1}$ at liquid helium temperatures and a tenfold increase in the resistivity of graphene. The irradiated devices remained latched in their high-resistivity state over many months. The ``on/off'' ratio of 10 for the resistivity of the photochemically-gated devices is similar to the best large-area single-layer graphene transistors demonstrated to date~\cite{Schwierz2010}. Very significantly, annealing the samples at 170$^\circ$C, just above the glass transition temperature of the polymers, reversed the effects of light and returned the graphene charge carrier density to its value prior to UV exposure. 

\begin{figure}
\centering
\includegraphics{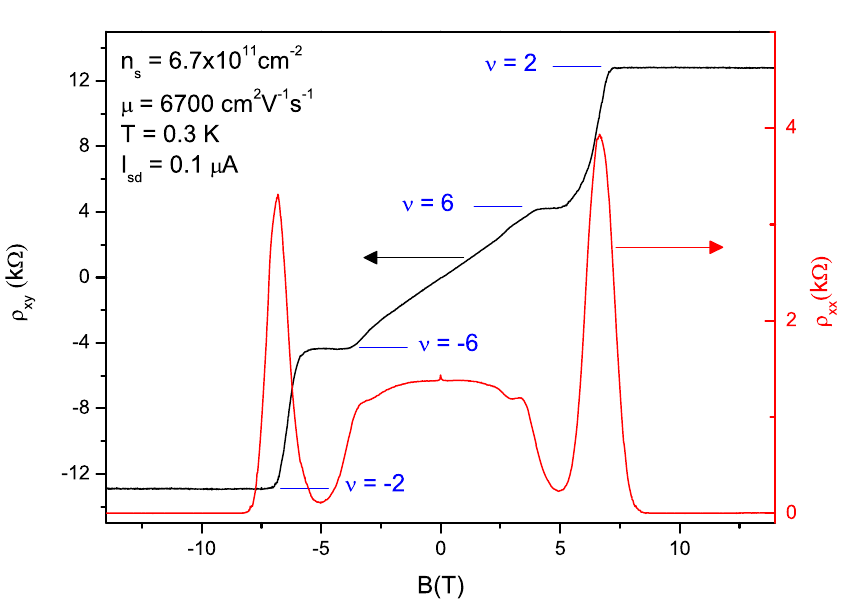}
\caption{\label{labelfig19} (Color online) Chemically gated graphene device. Transverse ($\rho_{xy}$) and longitudinal ($\rho_{xx}$) resistivity measurement. The horizontal lines indicate the exact quantum Hall resistivity values for filling factors $\nu = \pm 2$ and $\pm 6$.(Adapted from Ref.~\cite{Janssen2011a})}
\end{figure}

Figure \ref{labelfig19} shows the measurement of the longitudinal and transverse resistance as a function of magnetic field at low temperatures after encapsulation and UV illumination for the large Hall bar device ($160\ \rm\mu m\times 35\ \mu m$) with $n_s =6.7\times 10^{11}\ \rm cm^{-2}$and should be compared with fig.~\ref{labelfig13} where the carrier density was $8.5\times 10^{11}\ \rm cm^{-2}$. It is clear that covering the device with the polymer sandwich has dramatically improved the homogeneity of the device besides reducing the carrier density. 

\begin{figure}
\centering
\includegraphics[scale=0.65]{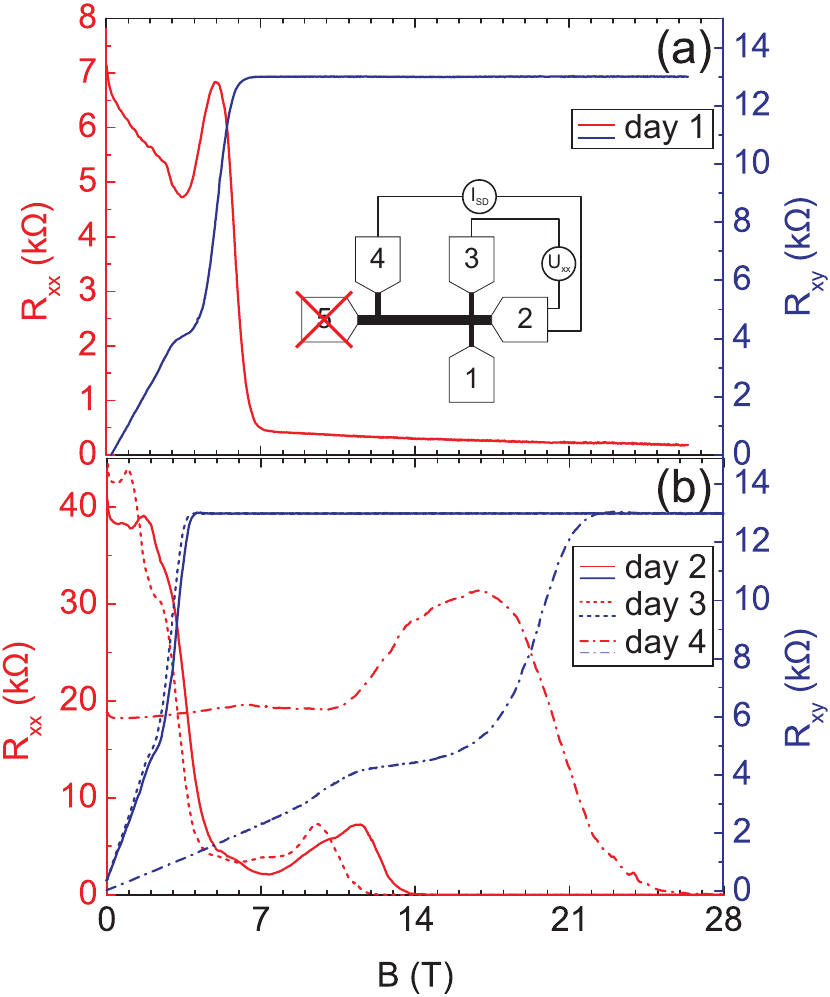}
\includegraphics[scale=0.8]{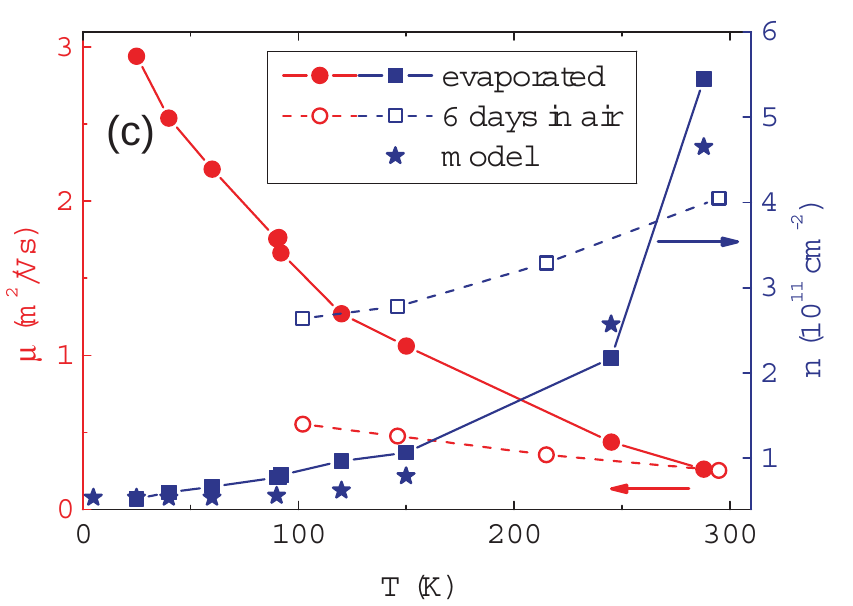}
\caption{\label{labelfig20} (Color online) (a) $R_{xx}$ and $R_{xy}$ at 4.2~K in a sample doped close to charge neutrality by F4-TCNQ. The data appear very similar to the photochemically gated device shown in fig.~\ref{labelfig19} except that $R_{xx}$ does not reach zero because one current contact was missing in this measurement. (b) Same as (a) but with contact repaired and measurements for successive days. On day 4 the chemical layer was rinsed off with water. (c) Carrier density and mobility for a sample close to the charge neutrality point as a function of temperature. The data for the device after six days in air demostrates the degradation of the F4-TCNQ layer. (From Jobst {\it et al.}~\cite{Jobst2010})}
\end{figure}

An alternative to photochemical gating is direct chemical gating as demonstrated by Jobst {\it et al.}~\cite{Jobst2010}. In this case the graphene is covered with tetrafluoro-tetracyanoquinodimethane (F4-TCNQ) molecules via thermal evaporation. Applying a monolayer of these acceptor-molecules reduced the carrier density down to $5.4\times10^{11}\ \rm cm^{-2}$ with an excellent mobility of $29000\ \rm cm^2V^{-1}s^{-1}$ [see fig.~\ref{labelfig20}(c)]. Although a good QHE was observed, the downside of this form of chemical gating is that it is not very stable in time, probably due to the absorption of water by the chemical layer [see fig.~\ref{labelfig20}(a) and (b)]. Recently, oxygen adsorption was used to reduce the carrier concentration in epitaxial graphene \cite{Pallecchi2012} and a quantum Hall effect has been observed at fields of $10-12\ \rm T$, yet, the issues with the sample stability have still to be solved. 

\section{Magnetic field dependent charge transfer}
\subsection{Filling factor pinning}
\begin{figure}
\centering
\includegraphics{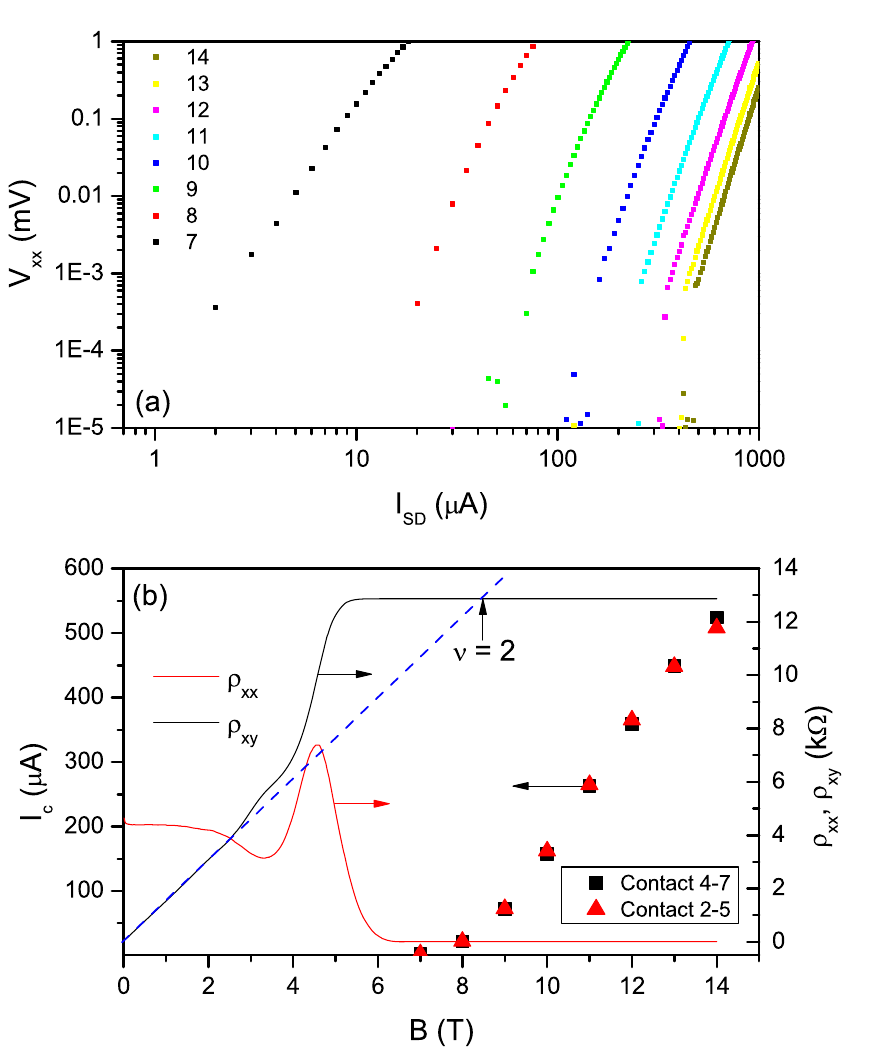}
\caption{\label{labelfig23} (Color online) (a) Measurement of $V_{xx}$ as a function of source-drain current at different values of magnetic field ranging from 7~T to (left hand curve) to 14~T (right curve) in steps of 1~T. (b) Transverse ($\rho_{xy}$) and longitudinal ($\rho_{xx}$) resistivity measurement at the reduced carrier density of $\rm 4.6\times 10^{11}\ cm^{-2}$ measured at $I_{sd}=1\ \rm\mu A$ together with the measured break-down current, $I_c$. Dashed blue line indicates position of the exact $\nu=2$ filling factor for the low field carrier density. (Adapted from Ref.~\cite{Janssen2012})}
\end{figure}

Figure~\ref{labelfig23}(b) shows magnetotransport data of a device in which the carrier density was reduced down to $\rm 4.6\times 10^{11}\ cm^{-2}$ by photochemical gating~\cite{Janssen2011a}. An extremely wide $\nu=2$ quantum Hall plateau can be seen and the $\nu=6$ plateau has all but disappeared in this case. A similar effect can also be seen in the low carrier density data of Ref. ~\cite{Jobst2010}. Janssen {\it et al.}~\cite{Janssen2011a} developed a model to explain this behaviour which is based on the pinning of the $\nu=2$ quantum Hall state driven by a magnetic field dependent charge transfer mechanism. They argue that, specific to graphene on SiC, the pinning of the $\nu=4N+2$ filling factors is determined by the dominance of the quantum capacitance~\cite{Luryi88}, over the classical capacitance, in the charge transfer between graphene and surface-donor states of SiC/G. 

\begin{figure}
\centering
\includegraphics{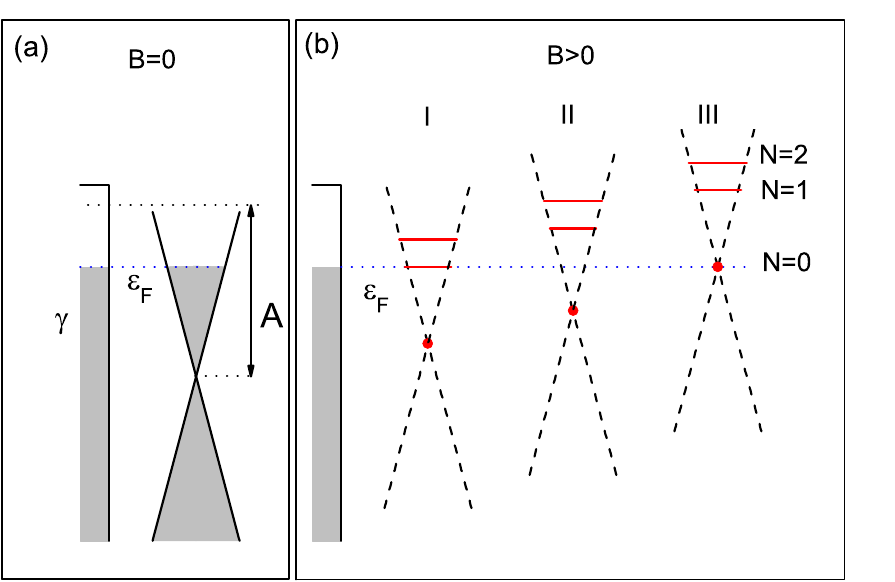}
\caption{\label{labelfig21} (Color online) Schematic band-structure for graphene on SiC in zero field (a) and in quantizing fields (b). (Adapted from Ref.~\cite{Janssen2011a})}
\end{figure}

The quantum capacitance of a two-dimensional electron system is the result of a low compressibility of the
electron liquid determined by the peaks in $\gamma_e$, the electronic density of states. For electrons in high-mobility GaAs/AlGaAs heterostructures in magnetic field, the quantum capacitance manifests itself in weak magneto-oscillations of the electron density~\cite{Eisenstein94,John04} due to the
suppressed density of states inside the inter-Landau level gaps. For epitaxial graphene on SiC, due to the short distance, $d\approx 0.3-0.4$~nm, between the buffer layer hosting the donors and graphene, the effect of quantum capacitance is much stronger, and the oscillations of electron density take the form of the robust pinning of the electron filling factor. Basically, the graphene layer and buffer layer exchange charge as the density of states oscillates with magnetic field. 

\begin{figure}
\centering
\includegraphics{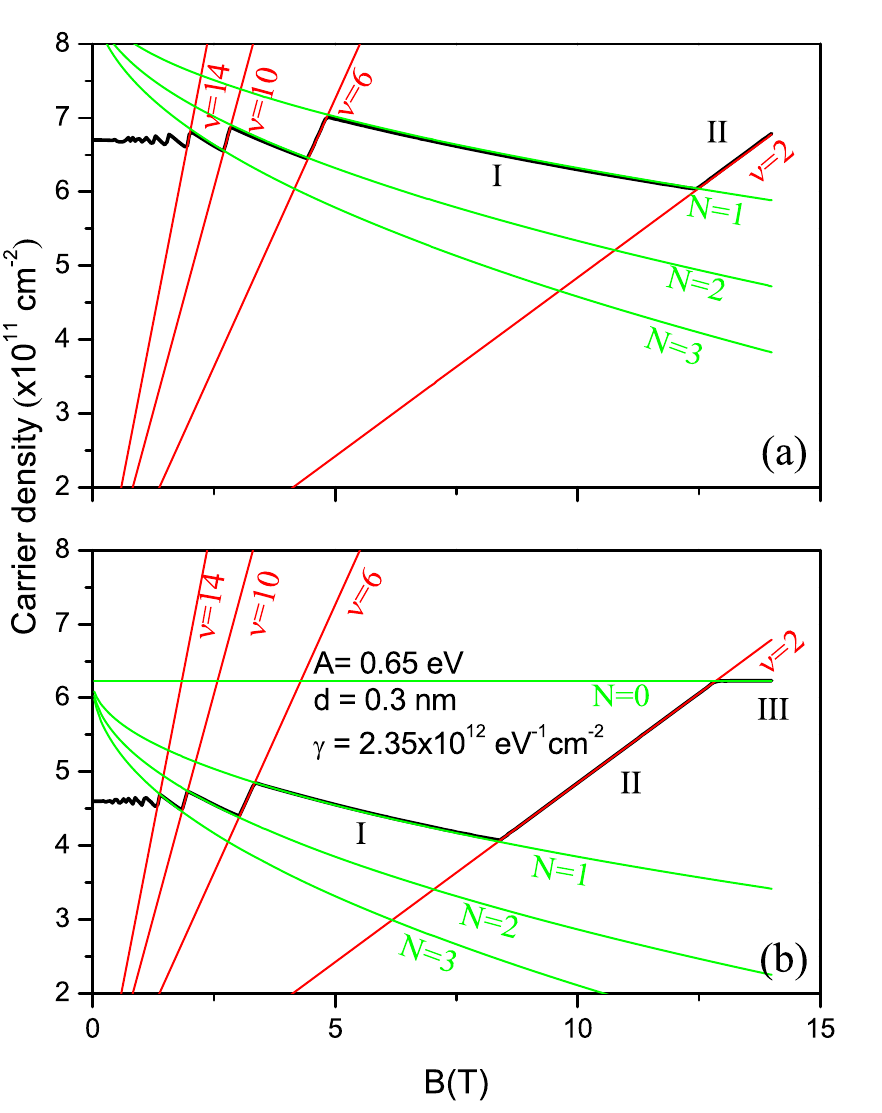}
\caption{\label{labelfig22} (Color online) Graphical solutions for the carrier density as a function of magnetic flux density, $n_s(B)$,
of the charge-transfer model given by Eq.~\ref{labeleq15} (black line) together with lines
of constant filling factor (red lines) and the magnetic field dependent carrier density in the Landau levels (green lines) for $n_g=5.4\times 10^{11}\rm\ cm^{-2}$ (a) and $n_g=8.1\times 10^{11}\rm\ cm^{-2}$ (b). (Adapted from Ref.~\cite{Janssen2011a})}
\end{figure}

The charge transfer in SiC/G is illustrated in the sketches in fig.~\ref{labelfig21}, for $B=0$ (a) and quantizing magnetic fields (b). In quantizing fields the Dirac cone transforms into discreet LL's which are depleted for successively larger magnetic fields. Again the transfer can be described using the charge balance Eq.~\ref{labeleq15} discussed earlier~\cite{Kopylov2010} and the magnetic field dependent solution is shown in fig.~\ref{labelfig22} for two different values $n_g$ (a higher $n_g$ results in a lower $n_s$). Interestingly, in regime II where the chemical potential in the system lies inside the gap between $N=0$ and $N=1$ LL, the carrier density increases linearly with the magnetic field, $n_s = 2eB/h$, due to the charge transfer from SiC surface. The final carrier concentration of graphene is determined mostly by the density of donor states $\gamma$ in the buffer layer and can be nearly 30\% higher than the zero-field density~\cite{Janssen2011a}.

Accurate quantum Hall resistance measurements require that the longitudinal voltage remains zero to ensure the device is in the non-dissipative state, which can be violated by the breakdown of the QHE at high source drain current levels. Fig.~\ref{labelfig23}(a) shows the determination of the breakdown current $I_{c}$ for different values of the magnetic field along the $\nu=2$ plateau. Here $I_{c}$ is defined as the source-drain current, $I_{sd}$, at which $V_{xx}\geq 10\ \rm nV$ [fig.~\ref{labelfig23}(b)].

From fig.~\ref{labelfig23}(b) we see that the breakdown current continues to increase in the $\nu=2$ quantum Hall state reaching $\rm \approx 500\ \mu A$ at the maximum field of 14~T. This behaviour is very different from that observed in conventional semiconductor systems where the breakdown current peaks at the exact integer filling factor~\cite{Jeckelmann2001} indicated by the dashed blue line in fig.~\ref{labelfig23}(b).  In graphene the filling factor is effectively pinned  at $\nu=2$ over a broad range of magnetic field resulting in a novel quantum Hall state which is ideally suited for high precision resistance metrology. The anomalous pinning is responsible for the extremely high breakdown current and wide operational parameter space of epitaxial graphene.

\begin{figure}
\centering
\includegraphics{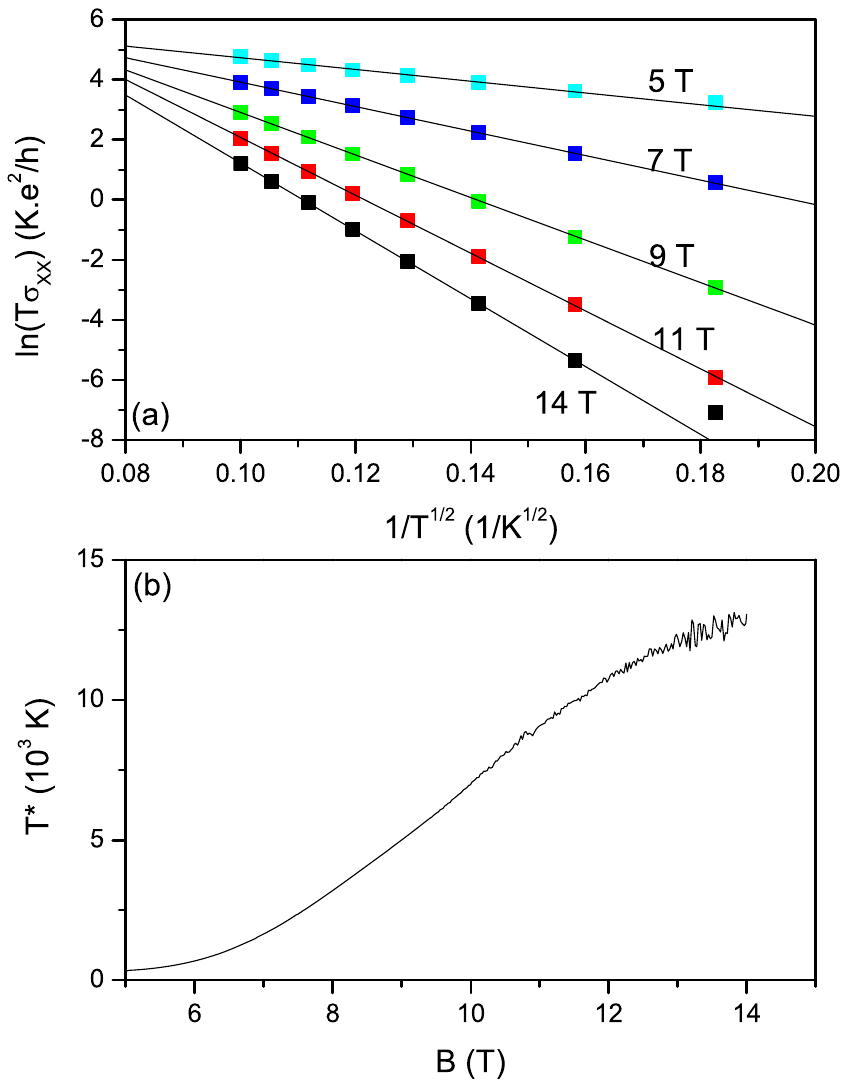}
\caption{\label{labelfig24} (Color online) (a) Variable-range hopping analysis of the data in fig.~\ref{labelfig23}(b) for the magnetic field range from 5 to 14~T and temperature range from 30 to 100~K. The axis are rescaled to give a straight line fit for $\sigma_{xx}(T)\propto (1/T)\exp{\left(-\sqrt{T^*/T}\right)}$. (b) Hopping temperature $T^*$ as function of magnetic field.}
\end{figure}

When the Fermi level lies in between the $N=0$ and $1$ Landau levels the activation energy for the dissipative transport $\hbar\sqrt{1/2} v_F/l_B$ is very large $\sim 1000\ \rm K$~\cite{Giesbers2007} (with $l_B$ is the magnetic length). For such a high activation energy, the low-temperature dissipative transport is most likely to proceed through the variable range hopping between surface  donors in SiC involving virtual occupancy of the Landau level states in graphene to which they are weakly coupled. Indeed, as shown in fig.~\ref{labelfig24}(a), the temperature dependence of the conductivity $\sigma_{xx}$ measured obeys an $\exp(-\sqrt{T^*/T})$ dependence typical of the hopping mechanism where $T^*$ is the defined as the hopping temperature. The $T^*$ values determined from the measurements at different magnetic fields are plotted in fig.~\ref{labelfig24}(b). The breakdown current rising with field to very large values [fig.~\ref{labelfig23}(a)] corresponds to  $T^*$ reaching extremely large values in excess of $10^4\ \rm K$ -- at least an order of magnitude larger than that observed in GaAs~\cite{Furlan1998} and more recently in exfoliated graphene~\cite{Giesbers2009,Bennaceur2010}. Such an extremely large hopping temperature is difficult to explain within the existing theoretical models and will need further investigation to fully understand.

\subsection{Robustness of the quantum Hall state}
\begin{figure}
\centering
\includegraphics{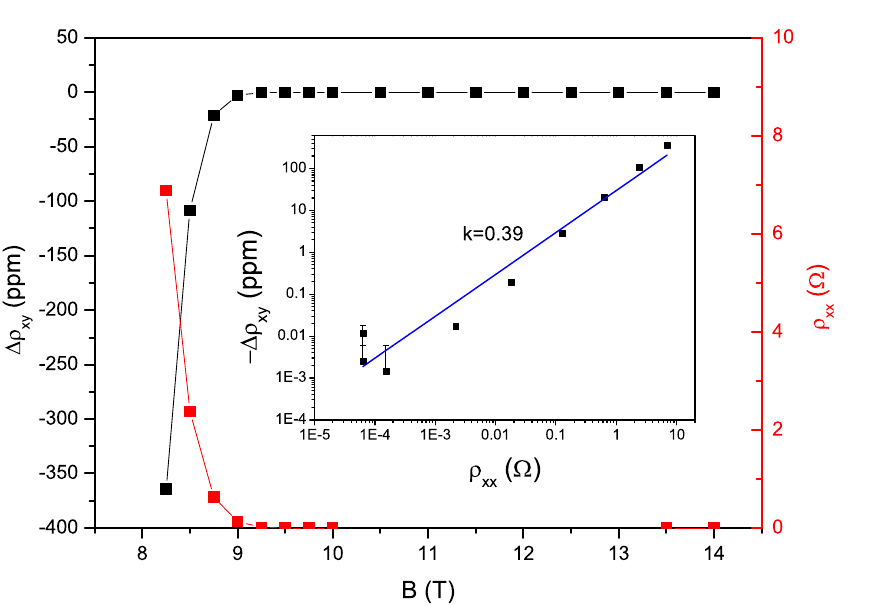}
\caption{\label{labelfig25} (Color online) Robustness of quantization as a function of magnetic field. The main panel shows $\Delta\rho_{xy}=2[\rho_{xy}^{\rm Graphene}(B)-\rho_{xy}^{\rm GaAs/AlGaAs}(B{\rm =10.5\ T)}]/R_{\rm K}$ and $\rho_{xx}$ as a function of magnetic flux density on the graphene device. The measurement current was $\rm 60\ \mu A$. $1 \sigma$-error bars for $\Delta\rho_{xy}$ are of the order of 5~ppb and not visible on the scale of this graph. The inset shows a plot of the same data but here $-\Delta\rho_{xy}$ is plotted as function of $\rho_{xx}$. The blue line is a linear fit. (Adapted from Ref.~\cite{Janssen2012})}
\end{figure}

Robustness of the quantum Hall state is important for resistance metrology and implies that the quantization is insensitive to variations in experimental parameters making it easy to realize the correct value of $h/\nu e^2$. Figure~\ref{labelfig25} demonstrates the robustness as a function of magnetic flux density for the $\nu=2$ quantum Hall state~\cite{Janssen2012}. Here $\Delta\rho_{xy}=2[\rho_{xy}^{\rm Graphene}(B)-\rho_{xy}^{\rm GaAs/AlGaAs}(B{\rm =10.5\ T})]/R_{\rm K}$ measures the difference between graphene and a fixed reference GaAs/AlGaAs device. The magnetic field for the GaAs/GaAlAs device is held constant at exactly $\nu=2$ and the magnetic field for the graphene sample is varied. In a separate measurement the longitudinal resistivity was measured for the graphene device and plotted in the same graph. The graph shows that the quantum Hall effect is accurate (of order 5~ppb) over a range of at least 4~T and only limited by the maximum magnetic field. The inset is a plot of $\Delta\rho_{xy}$ as a function of $\rho_{xx}$ and confirms the empirical relationship $\Delta\rho_{xy}=k\rho_{xx}$, here with $k=0.39$, which has been observed for traditional semiconductor systems~\cite{Jeckelmann2001} (The numerical value of $k$ is not universal and different for different contact pairs and different cool-downs of the device~\cite{Janssen2012}).

\begin{figure}
\centering
\includegraphics[scale=0.4]{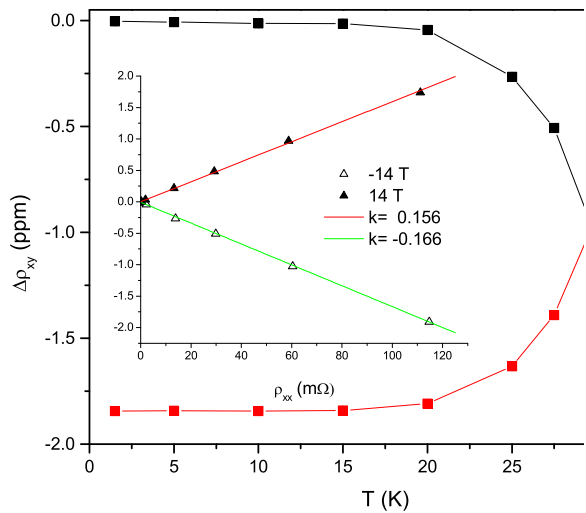}
\caption{\label{labelfig26} (Color online) Robustness of quantization as function of temperature as in fig.~\ref{labelfig25}. $B=14$~T for the graphene device and 10.5~T for the GaAs/AlGaAs device. The measurement current was again $\rm 60\ \mu A$. The inset shows a plot of the same data but here $\Delta\rho_{xy}$ is plotted as function of $\rho_{xx}$ for opposite magnetic field directions. Red and green lines are linear fits. (Adapted from Ref.~\cite{Janssen2012})}
\end{figure}

In figure~\ref{labelfig26} the same experiment is repeated but this time with temperature as the parameter at a magnetic field of 14~T. The graphene device shows good (ppb-level) quantization up to at least 15~K, however it is worth noting that the maximum magnetic field available is probably not yet at the plateau center and so the maximum operational temperature could be higher. The inset shows the plots of $\Delta\rho_{xy}$ versus $\rho_{xx}$ for opposite magnetic field directions from which one can determine that $k=0.16$. The sign of $k$ changes with the magnetic field direction which is again in accordance with that observed for semiconductor systems. A $k$-value of $0.16$ implies that a relative error in $R_{\nu=2}^{\rm Graphene}\leq$ parts in  $10^{10}$ is achievable when $\rho_{xx}\leq\ 10\ \mu\Omega$.

\section{Energy loss and breakdown}
The large breakdown currents observed in graphene mean there is significant heating of the electron gas in the quantum Hall state (for $100\ \rm \mu A$ this is more than $100\rm\ \mu W$ in a $35\ \rm \mu m$ wide channel). This in turn implies that the charge carriers in graphene must be very efficient in losing their energy to the lattice, a property which makes this material so attractive for modern electronics where the power densities are limiting performance. This property is somewhat counter intuitive given that the available phonon spectrum for scattering is very limited (up to room-temperature polar-optical phonon scattering and piezoelectric acoustic phonon scattering are negligible).

\begin{figure}
\centering
\includegraphics[scale=0.4]{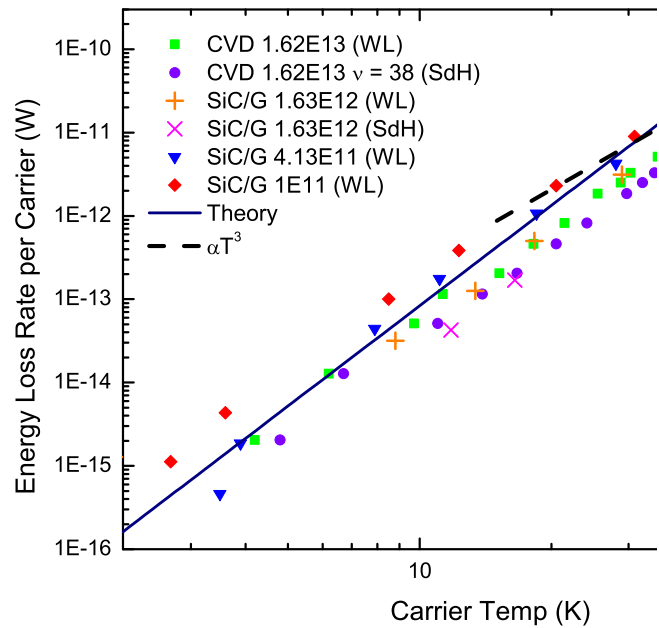}
\caption{\label{labelfig27} (Color online) Carrier energy loss rate as a function of electron temperature for a representative variety of samples and from the two techniques used. The data follow a similar trend for all samples, with the energy loss rate scaling by approximately $T^4$. An extrapolated $T^4$ dependence from low temperature calculations~\cite{Kubakaddi2009} is shown for a carrier density of $4.13\times 10^{11}\rm \  cm^{-2}$ and a $T^3$ dependence is also shown. (Adapted from Ref.~\cite{Baker2013})}
\end{figure}

Baker {\it et al.}~\cite{Baker2012,Baker2013} studied the energy loss rates of hot charge carriers in graphene produced by three different methods, exfoliated, epitaxial and CVD-grown samples. By measuring the amplitude of the Shubnikov-de Haas oscillations as a function of temperature and current, the carrier temperature can be determined as a function of current. For epitaxial graphene in which not enough SdH oscillations are present, the amplitude of the zero field weak localization peak was used to determine the carrier temperature. From these results the energy loss rates can be obtained as a function of carrier temperature which are shown in fig.~\ref{labelfig27}. 

The data in fig.~\ref{labelfig27} show a behaviour which is consistent with a $\sim T^4$ power law dependence. The solid line is based on a theoretical model by Kubakkadi for acoustic phonons, extrapolated from low temperatures~\cite{Kubakaddi2009}. Using this theoretical model, the electron energy relaxation time, $\tau_e$, can be obtained from the energy loss rate. Baker {\it et al.} typically find that $\tau_e$ in graphene is at least an order of magnitude shorter than that observed in GaAs over a wide temperature range.  This is despite the fact that optical phonon emission is making a strong contribution to the GaAs energy loss rate above $\sim50\ \rm K$. 

This high loss rate together with the large cyclotron energy explains the high breakdown current in graphene. This can be seen from the most commonly used model to predict the breakdown of QHE, which is the bootstrap-type electron heating model of Komiyama and Kawaguchi~\cite{Komiyama2000}. This model is based on the runaway heating which occurs when the quantum Hall effect begins to break down. Within this model, for graphene, the breakdown field $E_y$ is predicted to be

\begin{equation}
E_y=\sqrt{\frac{B\hbar\omega_c}{e\tau_e}}.
\end{equation} 

We see that with typical parameters for graphene ($\tau_e=4\ \rm ps$ and $\hbar\omega_c=126\ \rm meV$) we can expect a breakdown current 10 times higher than for GaAs ($\tau_e=100\ \rm ps$ and $\hbar\omega_c=17\ \rm meV$)~\cite{Baker2012}. These results suggest that the intrinsic properties of graphene make it a much better choice for the realization of a quantum resistance standard than traditional GaAs. 

\section{Universality of the QHE in epitaxial graphene}
As discussed in section~\ref{resistance-metrology}, in order to test the universality of the quantum Hall effect, two different QHE devices need to be set up at the same quantum Hall plateau so that a one-to-one comparison of  resistance can be made (in principle one could also compare different index plateaux if different winding ratios are available on the CCC). In 2010 NPL and the Bureau International des Poids et Mesures (BIPM) collaborated in a joint experiment to compare a graphene device and a GaAs/AlGaAs device directly against each other by using the BIPM traveling quantum Hall system~\cite{Janssen2011b}. 

\begin{figure}
\centering
\includegraphics{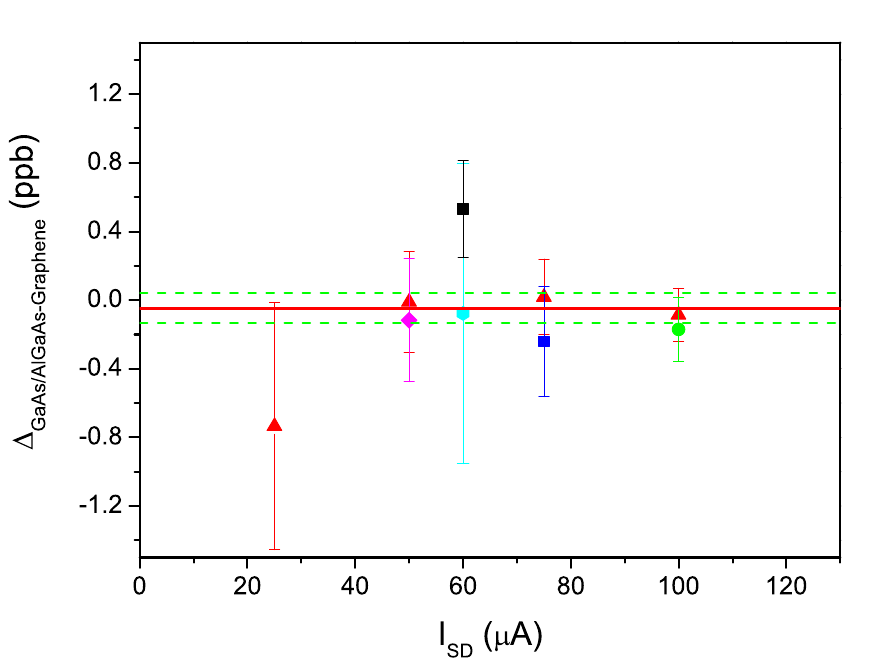}
\caption{\label{labelfig28} (Color online) Measurement of $\Delta_{\rm GaAs/AlGaAs-Graphene}$ expressed in parts per billion for the direct comparison of $R_{\rm H}^{\rm GaAs/AlGaAs}$ and $R_{\rm H}^{\rm Graphene}$ as a function of $I_{sd}$. The uncertainty bars represent the $\pm 1\sigma$ random error of the mean. Different symbols are explained in the text. The red line is the weighted mean of all the data points and the green lines signify $\pm 1\sigma$. (Adapted from Ref.~\cite{Janssen2011b})}
\end{figure}

The graphene sample was mounted in a 14~T/300~mK cryostat and connected to the Slave side of the CCC bridge. Two GaAs/AlGaAs samples were mounted in the transportable 11.7~T/1.2~K BIPM cryostat and connected to the Master side. The two GaAs samples used were traditional GaAs/AlGaAs heterostructures obtained from the PTB (device 1) and LEP (device 2). The red triangles in fig.~\ref{labelfig28} are the results for graphene against device 1 for 4 different source-drain currents in the devices. The pink diamond is a measurement for graphene against device 2 at at $I_{sd}=50\ \rm\mu A$. Here $\Delta_{\rm GaAs/AlGaAs-Graphene}=(R_{\rm H}^{\rm GaAs/AlGaAs}-R_{\rm H}^{\rm Graphene})/(R_{\rm K}/2)$ and each data point consists of an average of between 3 and 10 hours worth of data. The uncertainty increases for lower $I_{sd}$ because the signal-to-noise reduces for lower $I_{sd}$.

To check for errors due to non-zero $\rho_{xx}$ the measurements were repeated for non-opposite contacts (green dot and blue square). (Note that it is very difficult to measure $\rho_{xx}$ directly to the required level of precision.) Another test to check for small errors is to reverse the direction of magnetic field on the graphene sample. The results of this measurement is represented by the light blue hexagon. Finally, the devices were exchanged between the NPL and BIPM cryostats in order to eliminate small parasitic leakages (black square in fig.~\ref{labelfig28}). 

The data in fig.~\ref{labelfig28} show no sign of any systematic errors in the measurement campaign and so all results can be combined to give a weighted mean of $\Delta_{\rm GaAs/AlGaAs-Graphene}=(-4.7\pm 8.6) \times 10^{-11}$. The random noise of $8.6$ parts in $10^{11}$ dominates over the other components, estimated to have a combined standard uncertainty of $1.6$ parts in $10^{11}$~\cite{Janssen2012}. 

Previously our knowledge of the universality of the QHE has been limited to the level of 2 or $3\times\ 10^{-10}$ for comparisons between GaAs and Si or between identical GaAs devices~\cite{Hartland1991,Jeckelmann1997,Schopfer2007,Poirier2009}. However both GaAs and Si are traditional semiconductors with a parabolic bandstructure and governed by the same physics. Graphene is a semi-metal with a linear bandstructure and is described by Dirac-type massless charge carriers and so universality in terms of material independence goes well beyond the comparison between two semiconductors. The result does directly support the Thouless-Laughlin argument~\cite{Thouless1994} that the Hall conductivity is a topological invariant and is therefore a fundamental test of condensed matter theory. 

The value of $\Delta_{\rm GaAs/AlGaAs-Graphene}=(-4.7\pm 8.6) \times 10^{-11}$ on the material independence is the strongest evidence yet of the hypothesis that the resistance is quantized in units of $h/e^2$ is correct. This result underpins the assumptions made in the Watt balance experiment which links mechnanical and electrical units and thereby supports the pending re-definition of the SI-units for kilogram and ampere in terms of $h$ and $e$~\cite{Mills2006,Milton2010}. 

\section{The era of graphene-based metrology is only just beginning}

\subsection{Graphene single-electron pumps for better quantum current standards}
Single electron pumps are set to revolutionize electrical metrology by enabling the ampere to be re-defined in terms of the elementary charge of an electron \cite{Zimmerman2003}. Precise electrometers or high-value resistors can be calibrated directly using single electron pumps to a much higher precision than would be possible via a combination of Josepshon and quantum Hall effects~\cite{Giblin2012}. Pumps based on lithographically-fixed tunnel barriers in mesoscopic metallic systems \cite{Keller1996,Keller1999} and normal/superconducting hybrid turnstiles \cite{Pekola2008,Averin2008} can reach very small error rates, but only at MHz pumping speeds corresponding to small currents of the order $1\ \rm pA$. Tunable barrier pumps in semiconductor structures have been operated at GHz frequencies \cite{Fujiwara2008,Giblin2012}, but the theoretical treatment of the error rate is more complex and only approximate predictions are available \cite{Kashcheyevs2010}.

Connolly {\it et al.}~\cite{Connolly2013} have recently demonstrated a monolithic, fixed-barrier pump made entirely in graphene (see fig.~\ref{labelfig29}). Their adiabatic double quantum dot pump operates at frequencies up to 1.4~GHz with a predicted error rate as low as 0.01 parts per million at 90 MHz. Ten pumps operated in parallel would deliver 100 pA with metrological accuracy. The authors ascribed the high performance of the pump to a number of factors deriving from graphene's unique two-dimensional physical and electronic structure:

\begin{figure}[th]
  \centering
  \includegraphics[scale=0.6]{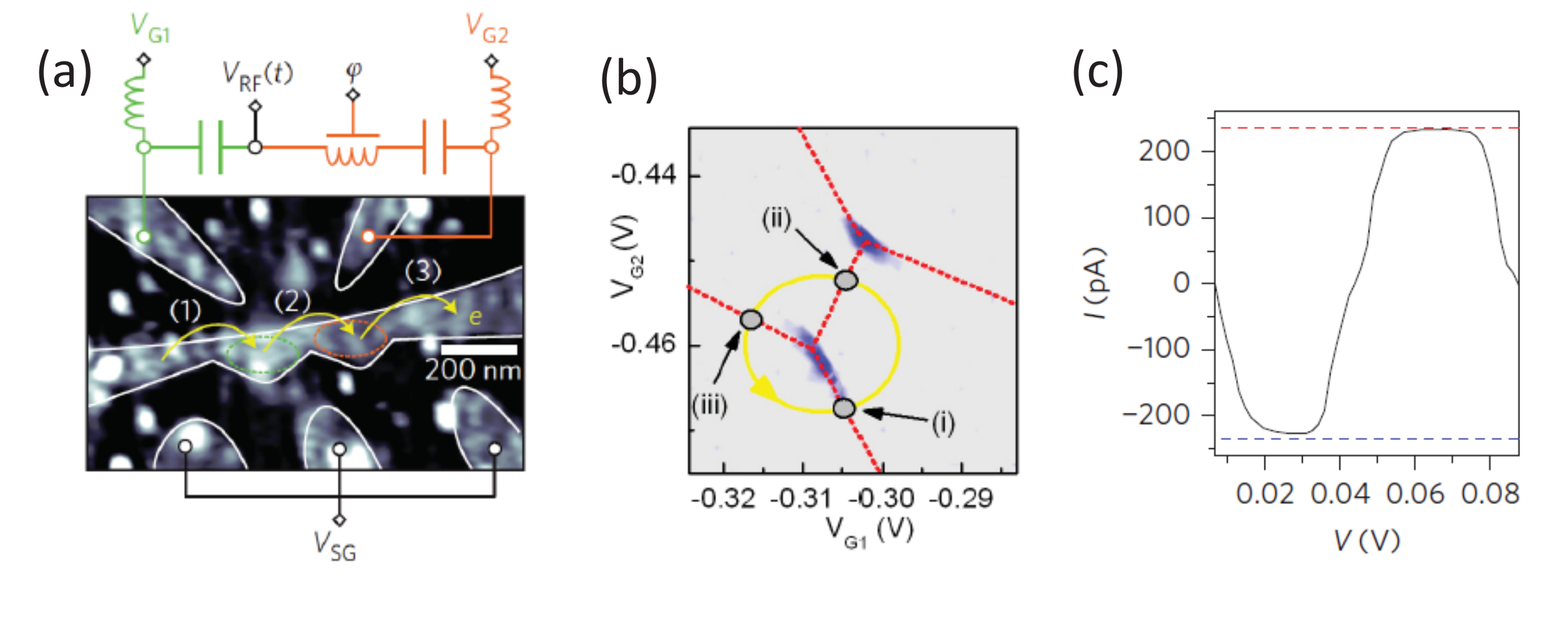}
  \caption{(Color online) Quantized charge pumping in graphene. (a) Atomic force micrograph of a double dot device (white lines indicate the edge of the graphene). An oscillating voltage $V_{RF}(t)$ is added to the d.c. voltages $V_{G1}$ and $V_{G2}$. (b) Source-drain current for small bias voltage as function of both gates. Red dashed lines indicate the edges of the stability diagram, revealing a honeycomb structure. A quantized current is pumped when the gate-voltage modulation encircles a triple point (yellow line)  (c) Plot of the pumped current along a line through two triple points. (From Connolly {\it et al.}~\cite{Connolly2013})}\label{labelfig29}
\end{figure}

Firstly, the presence of strong edge and potential disorder in lithographically defined graphene nanostructures leads to the formation of multiple quantum dots in the constrictions acting as tunnel barriers between the dots. Rather than this impeding pumping, the resulting capacitance and resistance of the random tunnel junctions between the quantum dots promotes a high intrinsic tunneling time while simultaneously suppressing co-tunneling events due to the large overall dissipation.

Secondly, the large interdot capacitance and linear electronic dispersion in graphene leads to a large and occupancy-dependent single-particle energy spacing near the Dirac point, $\Delta(N) = \hbar v_F / (d\sqrt{N})$ ($N \gg 1$ -- the number of electrons on the dot at a given Fermi energy, $d$ -- the dot diameter), which suppresses photon assisted interdot transitions and protects adiabaticity of the pump when operated at high frequency. This permits the interdot capacitance to be optimized through increased dot size, thereby improving the tolerances for parallelization, without compromising on accuracy.

Combined with the record-high accuracy of the quantum Hall effect and proximity induced Josephson junctions \cite{Jeong2011}, accurate quantized current generation brings an all-graphene closure of the quantum metrological triangle within reach \cite{Piquemal2000, Zimmerman2003}. In the metrological triangle experiment the resistance in terms of $h/e^2$, the voltage in terms of $h/2e$ and current in terms of $e$ are compared directly or indirectly against each other, revealing possible inconsistencies.

\subsection{Quantum Hall arrays for resistance metrology} 
The value of the quantized Hall resistance at $\nu=2$ of $\approx 12.906\ \rm k\Omega$ is not very practical for everyday resistance metrology which is concentrated around decade values of resistance. A large ratio, typically more than 100 to 1, has to be bridged in order to calibrate a $100\ \rm\Omega$ resistance which is a demanding measurement. A solution to this problem is the use of quantum Hall arrays with which a wide scale of resistance can be covered~\cite{Delahaye1993,Poirier2002}. The use of a quantum Hall arrays resistance standard (QHARS) has so far been limited. This is primarily due to low tolerance of Hall quantization in the semiconductor arrays to any spread of characteristics in individual devices leading to low overall breakdown currents and hence low signal-to-noise ratio. In addition, connecting several hundred Hall bars in a series or parallel network, which requires many bridging connections and all contact resistances to be low-ohmic, entails complex multilayer fabrication. In contrast, large area graphene offers a wide operational parameter space for the quantum Hall effect, high breakdown current and the ease with which low-ohmic contacts can be achieved. 

\begin{figure}[th]
  \centering
  \includegraphics[scale=0.6]{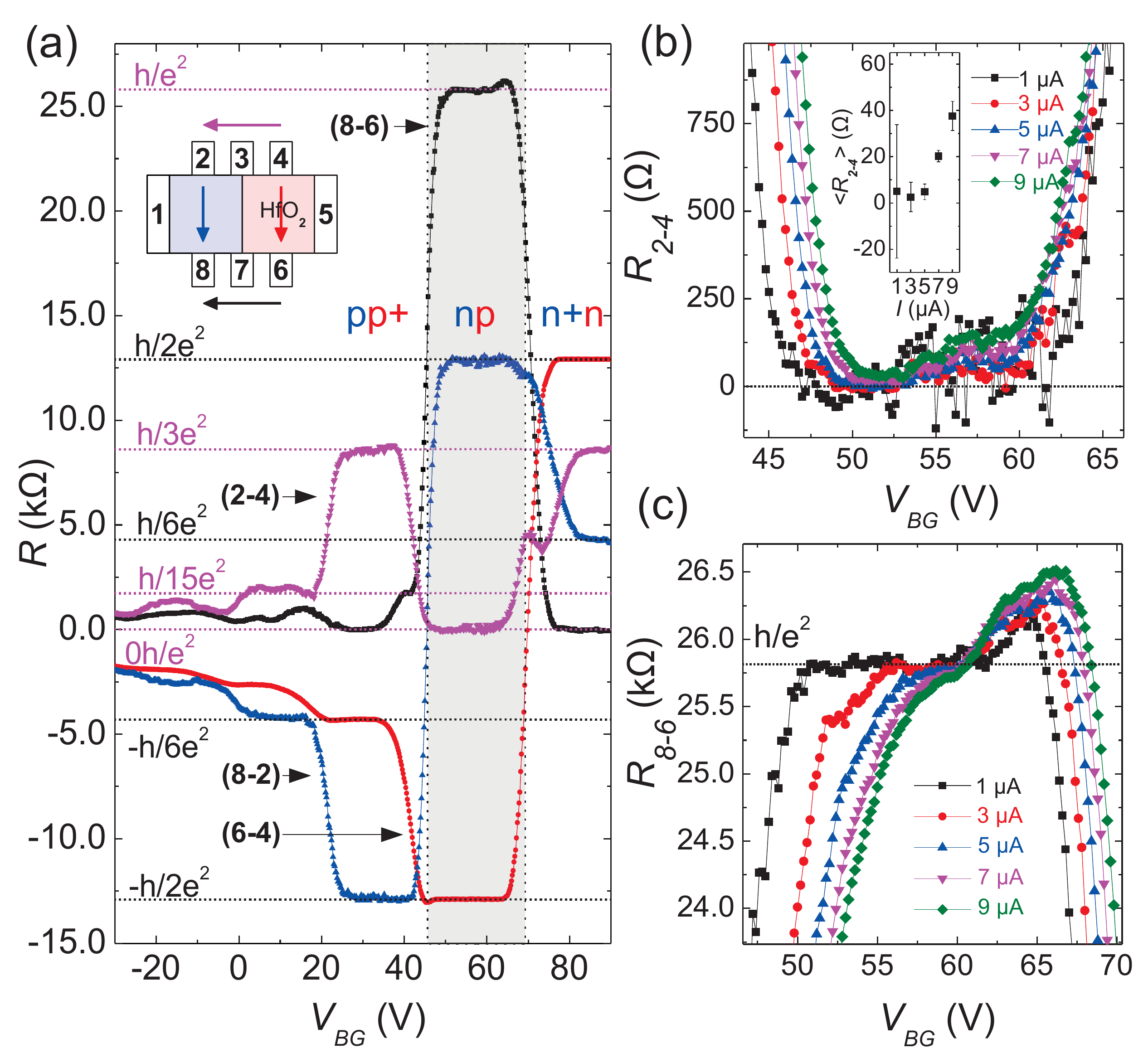}
  \caption{ (Color online) Graphene {\it p-n} junction array. (a) Four-terminal resistance measurement as a function of back-gate voltage and $1\ \rm\mu A$ source-drain current between contacts 1 and 5. Labels at curves denote the voltage contact pair. (b) Resistance measurement for different currents measured between contacts 2 and 4. (c) Same as (b) for contacts 6 and 8. (From Woszczyna {\it et al.}~\cite{Woszczyna2011b})}\label{labelfig30}
\end{figure}

In graphene the carrier type can be changed by application of a gate voltage which allows the realization of ambipolar structures in one device. Devices can be constructed with $p-n$ junctions or regions of different carrier density. Williams {\it et al.}~\cite{Williams2007} demonstrated that the magneto-transport in these devices is quantized in units of $h/e^2$ but with a new series of indices. This behaviour was explained by Abanin and Levitov~\cite{Abanin2007} as resulting from the equilibration of edge states along the junction interface. Subsequently, Woszczyna {\it et al.}~\cite{Woszczyna2011b} realized that this mechanism could be exploited to construct an array of quantum Hall devices without the need of bridging interconnects by creating alternate regions of $p$ and $n$ type domains [see fig.~\ref{labelfig30}(a)]. They measured vanishing longitudinal resistance and quantized Hall resistance in a two-segment array thereby demonstrating the feasibility of this novel approach [see fig.~\ref{labelfig30}(b) and (c)]. 

\subsection{Constant light absorption in a broad spectral range for radiometry}
Graphene is an attractive material for use in optical detectors because it absorbs light from mid-infrared to ultraviolet wavelengths with nearly equal strength, $\approx \pi \alpha$ per layer~\cite{Nair2008} with $\alpha$ the fine-structure constant. Because of its small heat capacity, very weak coupling to thermal baths and rapid thermalization~\cite{Betz2012,Baker2013} it is particularly well suited for use in cooled bolometers - devices that measure the power of incident electromagnetic radiation via the heating of material with a temperature-dependent electrical resistance. Jun Yan {\it et al.}~\cite{Yan2012} have demonstrated a graphene bolometer, which exhibits a noise-equivalent power $33~\rm fW~Hz^{-1/2}$ at $5~\rm K$ that is several times lower, and intrinsic speed $>1~\rm GHz$ at $10~\rm K$ that is three to five orders of magnitude higher than commercial silicon bolometers and superconducting transition-edge sensors at similar temperatures. Making use of wide-bandwidth noise thermometry, Fong {\it et al.}~\cite{Fong2012} were able to measure temperature oscillations with a period of $430~\rm ps$.
Modern absolute radiometers~\cite{Martin1985}, devices that measure the radiant power of electromagnetic radiation, are used to realise the candela -- the SI unit of luminous intensity. They are based on the substitution principle, whereby the power of the radiant heating of an absorbing material is compared to the electrical heating, and operate at cryogenic temperatures. Although the absorptivity can reach $99.998 \%$ it is never exactly known leading to uncertainties in the measurements. Currently primary standard cryogenic radiometers can measure the power of an intensity stabilized laser beam to about $0.01\%$. Combination of the accurately known (though relatively small) wide-band optical absoption in graphene and high sensitivity and speed of graphene bolometers, can lead to novel quantum-based, fast, fairly compact and simple radiometers, linking the optical scale to the fine structure constant.

\subsection{Exceptional mechanical properties for displacement, force and mass sensors on the nanoscale}

Although the most precise mass sensing experiments with a resolution of 1.7 yoctogram ($1~{\rm yg}=10^{-24}~{\rm g}$), which corresponds to the mass of one proton, have been performed using carbon nanotube NEMS \cite{Chaste2012}, replacing them with graphene \cite{Bunch2007} brings the advantage of a larger area for force application and better coupling to the displacement sensor, although compromising the resonant frequency and hence the quality factor. The fundamental problem of the resonance-based mass (or force) measurements is that one cannot tell the mass (force) distribution on the resonator from the frequency shifts (one cannot hear the shape of the drum \cite{Gordon1992}) even when multi-mode measurements are carried out. In the case of mass sensing, this problem can potentially be solved by functionalizing graphene or nanotube, so that the rest mass (e.g., a gas molecule) is absorbed at an atomically defined position. For force sensing graphene NEMS can be driven by the local radiation pressure and read-out also optically. Having in mind the quantized optical absorption of graphene this may give a unique way of realizing sensitive force measurements linked to fundamental constants. In addition, graphene strain gauges with either electrical or optical readouts look very promising. Graphene is the only crystal which can be stretched by $20\%$ \cite{Lee2008}, thus enhancing the working range of such sensors significantly.

\subsection{Extreme sensitivity of electron transport for bio-molecular and gas metrology}
Detection of individual events when a gas molecule attaches to or detaches from graphene's surface have been demonstrated \cite{Schedin2007}. The adsorbed molecules change the local carrier concentration in graphene one by one electron, which leads to step-like changes in resistance. The achieved sensitivity is due to the fact that graphene is an exceptionally low-noise material electronically. This opens up unprecedented applications requiring counting of species, such as the measurements of the amount of substance on the nanoscale. For analytical applications however these measurements should be made discriminatively between different atomic/molecular entities, which requires chemical functionalization of the graphene sensor.

We can confidently predict that other, even more exciting applications of graphene in metrology will be investigated. It can be argued \cite{Novoselov2012} that the major advantage of graphene sensors is their multi-functionality. A single device can be used in multidimensional measurements (for example, strain, gas environment, pressure and magnetic field). Whether or not graphene will fulfill the expectations is yet to be seen. But if we can practically realize the potential of this unique two-dimensional material, than the impact on technology and metrology will no doubt be revolutionary.

\section{Conclusions}
The journey from original discovery of the QHE in graphene to a quantum resistance standard which outperforms the established technology in many aspects has been remarkably short. For epitaxial graphene the robustness of the quantization in terms of temperature, magnetic field and source-drain current is exceptional. The material is cheap and relatively easy to fabricate and process. It allows for the realization of a quantum resistance standard with modest means, e.g. a small superconducting magnet and cryocooler. As such it will improve the proliferation of quantum standards and allow many smaller laboratories to realize their own resistance scale. One even could envisage university students being able to perform QHE experiments, much in the same way as the discovery of High-$T_c$ superconductors enabled table-top experiments with Josephson junctions in many science classes. 

\ack
We would like to thank our collaborators Rositza Yakimova, Sergey Kopylov, Olga Kazakova, Nick Fletcher, Roland Goebel, Jonathan Williams, Dale Henderson, Stephen Giblin, Pravin Patel, John Gallop, Thomas Bj{\o}rnholm, Kasper Moth-Poulsen, Karin Cedergren, and Mikael Syv{\"a}j{\"a}rvi. We also like to thank Thomas Seyller, Tord Claeson, Klaus von Klitzing, Konstantin Novoselov and Andre Geim for many illuminating discussions.  This work was supported by the NMS Pathfinder, IRD and EMT Programmes, Swedish Research Council and Foundation for Strategic Research, EU FP7 STREPs ConceptGraphene and SINGLE, EPSRC grant EP/G041954 and the Science \& Innovation Award EP/G014787.


\end{document}